\documentclass[journal]{IEEEtran}

\ifCLASSINFOpdf
\else
\fi
\usepackage{graphicx}
\usepackage{float}
\usepackage{amsmath}
\usepackage{amssymb}
\usepackage[utf8]{inputenc}
\usepackage{caption}
\usepackage{cite}
\usepackage{lipsum}
\usepackage{multirow}
\usepackage{array}
\usepackage{enumitem}
\usepackage{geometry}
\usepackage{caption}
\usepackage[export]{adjustbox}
\usepackage{subfig}
\usepackage[usenames,dvipsnames]{xcolor}
\usepackage{tcolorbox}
\usepackage{tabularx}
\usepackage{array}
\usepackage{booktabs}
\usepackage{colortbl}
\usepackage{subfig}
\tcbuselibrary{skins}

\newcolumntype{Y}{>{\centering\arraybackslash}X}

\tcbset{tab1/.style={fonttitle=\bfseries\large,fontupper=\normalsize\sffamily,
colback=yellow!10!white,colframe=red!75!black,colbacktitle=Salmon!40!white,
coltitle=black,center title,freelance,frame code={
\foreach \n in {north east,north west,south east,south west}
{\path [fill=red!75!black] (interior.\n) circle (3mm); };},}}

\tcbset{tab2/.style={enhanced,fonttitle=\bfseries,fontupper=\normalsize\sffamily,
colback=yellow!10!white,colframe=red!50!black,colbacktitle=Salmon!40!white,
coltitle=black,center title}}
\usepackage{array}
\newcolumntype{P}[1]{>{\centering\arraybackslash}m{#1}}
\newcolumntype{N}{@{}m{0pt}@{}}

\usepackage{hyperref}
\begin{document}
%
\title{Nanoantennas and Nanoradars: The Future of Integrated Sensing and Communication at the Nanoscale

}



%

\author{M. Javad~Fakhimi,~\IEEEmembership{Student Member,~IEEE,}
        Ozgur~B. Akan,~\IEEEmembership{Fellow,~IEEE}
\thanks{The authors are with the Center for neXt-generation Communications (CXC), Department of Electrical and Electronics Engineering, Ko\c{c} University, Istanbul 34450, Turkey (e-mail: \{mfakhimi22, akan\}@ku.edu.tr).
\par
O. B. Akan is also with the Internet of Everything (IoE) Group, Electrical Engineering Division, Department of Engineering, University of Cambridge, Cambridge CB3 0FA, UK (email: oba21@cam.ac.uk).
\par
This work was supported by the AXA Research Fund (AXA Chair for
Internet of Everything at Ko\c{c} University).}}

\maketitle

\begin{abstract}
Nanoantennas, operating at optical frequencies, are a transformative technology with broad applications in 6G wireless communication, IoT, smart cities, healthcare, and medical imaging. This paper explores their fundamental aspects, applications, and advancements, aiming for a comprehensive understanding of their potential in various applications. It begins by investigating macroscopic and microscopic Maxwell's equations governing electromagnetic wave propagation at different scales. The study emphasizes the critical role of Surface Plasmon Polariton (SPP) wave propagation in enhancing light-matter interactions, contributing to high data rates, and enabling miniaturization. Additionally, it explores using two-dimensional materials like graphene for enhanced control in terahertz communication and sensing. The paper also introduces the employment of nanoantennas as the main building blocks of Nano-scale Radar (NR) systems for the first time in the literature. NRs, integrated with communication signals, promise accurate radar sensing for nanoparticles inside a nano-channel, making them a potential future application in integrated sensing and communication (ISAC) systems. These nano-scale radar systems detect and extract physical or electrical properties of nanoparticles through transmitting, receiving, and processing electromagnetic waves at ultra-high frequencies in the optical range. This task requires nanoantennas as transmitters/receivers/transceivers, sharing the same frequency band and hardware for high-performance sensing and resolution.
\end{abstract}

%
\IEEEpeerreviewmaketitle

\section{Introduction}
\IEEEPARstart{R} apid advancement of wireless communication technologies has changed the way we access information in our increasingly connected world. As we approach the era of 6G wireless communication, the demand for higher data rates, lower latency, and enhanced connectivity continues to dominate. To address these challenges, researchers are exploring new technologies that can reveal the unique properties of terahertz frequencies (0.1 to 10 THz) \cite{Intro _ ref 1}. Among these transformative technologies, nanoantennas are proposed as promising and practical structures with the potential to reshape wireless communication, revolutionize the Internet of Things (IoT) and healthcare, and advance medical imaging \cite{Intro _ ref 2}.
These nanoscale structures enable hyper data transmission rates and high capacity, making them ideal for addressing the burgeoning data demands of the 6G era \cite{Intro _ ref 4}. Furthermore, the development of nanoantennas as a crucial building block of Nano-scale Radar systems (NRs) is critical. Nanoantennas play a significant role in signal transmission and reception, enabling the tracking of spatial and physical/electrical properties of nanoparticles (targets) in nano-channels. Moreover, the use of nanoantennas in the field of medicine has demonstrated exceptional potential in diverse medical applications including imaging, biosensing, disease detection, drug delivery, photodynamic therapy (PDT), real-time monitoring, and photothermal therapy \cite{Medical 1,Medical 2,Medical 3,Medical 4,Medical 5,Medical 6,Medical 7,Medical 8,Medical 9,Medical 10}.
The foundation of nanoantennas lies in the principles of electromagnetic theory, expressed through the macroscopic and microscopic Maxwell's equations \cite{Marder book,Maxwell in Nanoscale,Maxwell's Treatise,Jackson 1999 book}. These fundamental equations explain the propagation of electromagnetic waves, including those at terahertz frequencies and optical frequencies, providing insights into the intricate behavior of nanoantennas and their interactions with electromagnetic radiation \cite{Intro _ ref 6,Plasmonic Book 1}.
 Surface Plasmon Polaritons (SPPs) emerge as a pivotal aspect of nanoantenna research, facilitating enhanced light-matter interactions and confinement of terahertz waves \cite{Nanoantenna _ Main Book} \cite{SPP_first,SPP_second,SPP_third}. SPP wave propagation enables significant control over electromagnetic fields, contributing to the high data rates and miniaturization capabilities of nanoantennas \cite{SPP _ ref 1}.
  Furthermore, the quantum mechanical perspective of nanoantennas offers a unique insight into the behavior of these structures in the microscopic world \cite{Intro _ ref 8}, \cite{Plasmonic Book 1,Plasmonic Effect _ Review,Quantum Mechanics _ Book,Quantum Plasmonics _ Paper 2,Quantum Plasmonics_Paper 1,Quantum_first}. The coexistence of classical and quantum phenomena in nanoantennas manifest new paths for quantum communication and sensing, offering a paradigm shift in future applications \cite{Intro _ ref 9}. Quantum mechanics provides a deeper understanding of the underlying mesoscopic and microscopic physics, and helps us to explain the fundamentals of the interaction between the light and the matter, which is considered to be the root of differences between the classical antenna theory, and the nanoantenna radiation rules.
   Researchers have harnessed these principles to fabricate nanoantennas that efficiently receive and transmit data while minimizing energy consumption and space requirements \cite{Intro _ ref 10}. Additionally, the utilization of two-dimensional materials, such as graphene, for manufacturing more efficient and more compact structures to fit in the nanoscale, has been investigated in this paper \cite{Intro _ ref 11}. 
   To realize the full potential of nanoantennas in real-world applications, researchers have focused on the fabrication and characterization methods of these nanoscale structures \cite{Intro _ ref 12}. Fabrication techniques at the nanoscale, such as Photolithography \cite{Fabrication _ 1}, Electron Beam Lithography (EBL) \cite{Fabrication _ 2,Fabrication _ 3}, Focused Ion Beam Lithography (FIBL) \cite{Fabrication _ 4,Fabrication _ 5}, Nanoimprint Lithography (NIL) \cite{Fabrication _ 6}, Roll-to-Roll Printing \cite{Fabrication _ 7}, and Solid-state Superionic Stamping \cite{Fabrication _ 8} allow precise construction of nanoantennas with subwavelength dimensions, making them suitable for integration into highly compact devices and systems. Characterization methods, such as Optical Microscopy (OM) \cite{characterization _ 1}, Scanning Electron Microscopy (SEM) \cite{characterization _ 2}, Scanning Tunneling Microscopy (STM) \cite{characterization _ 3}, Transmission Electron Microscopy (TEM) \cite{characterization _ 4}, and Atomic Force Microscopy (AFM) \cite{characterization _ 5} enable engineers to analyze the performance and functionality of nanoantennas under various conditions, providing valuable insights for optimization and application-specific customization.

This paper aims to underscore the significance of ongoing research and progress in the realm of nanoantennas, emphasizing on their design, manufacturing considerations, and foundational elements. Meanwhile, the concept of nanoscale radar systems has been proposed for the first time. Considering numerous emerging applications of nanoantennas and nanoradars in future applications across diverse domains, notably including 6G wireless communications and Internet of Things (IoT), and Medical applications, this work tries to encourage researchers and  come up with a starting point in further work and studies.\\
The rest of the paper is organized as follows. The necessary concepts and subjects to fully-understand the functionality of nanoantennas and nanoradars, such as the governing radiation rules at nanoscale, as well as the SPP waves propagation and excitation methodologies, are given in Sec. \ref{Background}. Then, the principles and parameters of nanoantenna theory, and their applications, such as in NR systems are discussed in Sec. \ref{THREE} and in Sec. \ref{FOUR}, respectively. At last, Sec. \ref{FIVE} contains the conclusion and future research direction for nanoradar systems.


\section{Background Knowledge} \label{Background}
\subsection{Antenna Theory}

   \subsubsection{Radiation in Macroscopic Range}  According to the well-established principles of electromagnetism, antennas function by radiating electromagnetic waves when influenced by sources like electric and magnetic charges, as well as current densities \cite{Electrodynamics Griphits Book}.
    To generate electromagnetic waves, oscillating fields in time are required. By designing the antenna structure appropriately, these waves can be guided along a transmission line (the body of the antenna) before being emitted into free space.
     When studying an antenna system, it is essential to identify the sources of radiation and calculate the resulting electromagnetic fields using relevant equations. This analysis helps us understand how the antenna radiates energy into the surrounding space. Once the radiation has been determined, we can evaluate the antenna's performance using commonly used metrics such as gain, directivity, and radiation efficiency. These measures provide insights into the antenna's effectiveness in transmitting or receiving electromagnetic signals.\cite{Balanis Book,Balanis Paper}.

    Maxwell's equations, the fundamental equations of electromagnetism, have been widely studied and applied in the \textit{low-frequency regime}, which includes microwave and infrared frequencies. At these frequencies, materials such as metals with high conductivities inhibit the propagation of electromagnetic waves within them. As a result, in microwave applications, metals are often approximated as Perfect Electric Conductors (PECs) to simplify the analysis.
       Generally speaking, Maxwell's equations can solve all electromagnetic boundary value problems, including antenna configurations. Assuming an $e^{j\omega t}$ time convention, in a lossless medium with $\sigma = 0$, we may recap the equations as follows:
       \begin{align}
         &\nabla \cdot \textbf{D}  = \rho_{e} \label{Maxwell first} ,\\
         &\nabla \cdot \textbf{B} = \rho_{m} , \\
         &\nabla \times \textbf{E} = -\textbf{M}-j\omega \mu \textbf{H} , \\
         &\nabla \times \textbf{H} = +\textbf{J} + j\omega \varepsilon \textbf{E} , \label{Maxwell last}
       \end{align}
       where \textbf{E}, \textbf{D}, \textbf{B}, \textbf{H}, and \textbf{J} are the electric field, electric displacement field, magnetic flux, magnetic field, and electric current density, respectively, while $\rho_{e}$, and $\rho_{m}$ denote the electric and magnetic charges. In these equations, $\textbf{J}$, $\textbf{M}$, $\rho_{e}$, and $\rho_{m}$ can be considered as the sources of radiation. Also, the Ohm's Law can be restated as $\nabla \cdot \textbf{J} = -j\omega \frac{\rho_{e}}{\varepsilon}$, and $\nabla \cdot \textbf{M} = -j\omega \frac{\rho_{m}}{\mu}$ accordingly. Naturally $\rho_{m}=0$, and the magnetic flux $\textbf{B}$ is solenoidal, which implies $\nabla \cdot \textbf{B} = \nabla \cdot \textbf{H} =0$. 
       Inspired by these equations, we may now investigate the microscopic regime, to realize the rules of propagation at nano scales.
        \subsubsection{Nanoscale Radiation}  As observed in the preceding section, Maxwell's equations exhibit no dependency on the operating frequency. This implies that the relationships between electromagnetic fields remain consistent throughout the entire spectrum.
         However, the behavior of propagating waves changes when we increase the frequency to the Terahertz gap and above. Three fundamental questions arise in this case: First, what exactly happens to Maxwell's equations at such high frequencies? Do they remain unchanged, or do modifications need to be made to account for the new effects and phenomena? Second, which materials can tolerate these high-frequency radiations, and which substances are suitable for constructing antennas to operate in this regime? Third, how do the structure and propagation environment of the waves change at these frequencies? Are there new behaviors, limitations, or advantages brought about by the higher frequency range?\\
           First of all, let's examine the wave equation derived from Maxwell's equations. The radiated electric and magnetic fields can be expressed by applying the curl operation to one of the curl equations of Maxwell's equations and substituting the resulting expression into the other equation. By doing this, we obtain
           \begin{align}\label{Wave-equation}
             \nabla^{2}\textbf{E}-\mu \varepsilon \frac{\partial^{2}\textbf{E}}{\partial t^{2}}= \nabla (\nabla \cdot \textbf{E})+\mu \frac{\partial \textbf{J}}{\partial t} ,
           \end{align}
           where the RHS represents how the sources are participating in radiation. Now if we consider a source-free region with a uniform electric charge density ($\nabla \cdot E=0$), we obtain
           \begin{align}\label{Wave-equation source-free}
             \nabla^{2}\textbf{E}-\mu \varepsilon \frac{\partial^{2}\textbf{E}}{\partial t^{2}}=0 .
           \end{align}
            We have assumed constant values for permittivity and permeability, which holds true in a homogeneous and isotropic environment. However, at the nanoscale, the assumptions of homogeneity and isotropy may not be valid due to the significant influence of quantum effects. In contrast to macroscopic regions, where microwaves belong, the nanoscale is characterized by the presence of quantum phenomena such as quantization, wave-particle duality, plasmonic effect, uncertainty principle, and superposition \cite{Quantum_first}. These effects play critical roles, and waves can interact with matter at molecular scales. One notable quantum effect in nanoscale radiations is the plasmonic effect, which gives rise to the generation of plasmons. Plasmons confine the electromagnetic waves near the surface of a metal, leading to the formation of Surface Plasmon Polaritons (SPP) or Localized Surface Plasmons (LSP) \cite{SPP_first,SPP_second}. SPPs are propagating and dispersive electromagnetic waves that are coupled to the electron plasma of a conductor at a dielectric interface. They exhibit unique properties and allow for the confinement and manipulation of light at the nanoscale. On the other hand, LSPs are non-propagating excitations of the conduction electrons in metallic nanostructures, coupled to the electromagnetic field. LSPs are responsible for enhanced light-matter interactions and localized field enhancements. These two types of electromagnetic waves are explained in more details in the next section.
            Since quantum effects play a critical role at the nanoscale, they directly influence the electrical and magnetic properties of the medium, including its permittivity and permeability. This can lead to the medium becoming anisotropic (with properties dependent on direction) or dispersive (with properties dependent on frequency). To accurately describe these effects, we represent permittivity and permeability as complex tensors in their most general forms. Consequently, we obtain
            \begin{align}\label{general form of wave equation}
              \nabla \times \nabla \times \textbf{E} -\mu_{0}\varepsilon_{0}\omega^{2}\underline{\mu}\underline{\varepsilon}\textbf{E}=0 .
            \end{align}
            In comparison to the conventional wave equation for macroscale radiations, the general form of the wave equation (\ref{general form of wave equation}) in the nanoscale is often more complex and challenging to solve. In many cases, analytical solutions for this equation may not be readily available. Consequently, numerical methods, such as the Finite-Difference Time-Domain (FDTD) and Finite Element Method (FEM), are commonly employed to analyze the behavior of nanoscale radiations in various geometries and structures. The FDTD method discretizes time and space, allowing for the numerical approximation of the wave equation. By updating the fields over small time steps and spatial grid points, the FDTD method can simulate the propagation of electromagnetic waves and their interaction with nanoscale structures. This method is widely used for its simplicity and ability to handle a wide range of structures and geometries. On the other hand, the Finite Element Method (FEM) approximates the solution of the wave equation by dividing the domain into smaller elements. This numerical method is particularly suitable for irregular geometries and complex material properties. By solving the equation within each element and considering the interactions between adjacent elements, the FEM can accurately model the behavior of nanoscale radiations. Both FDTD and FEM, along with other numerical methods, have become indispensable tools for analyzing and understanding the behavior of nanoscale radiations, as they provide computational solutions to the complex wave equations that describe these phenomena \cite{Balanis Paper,Maxwell in Nanoscale}. In the following section, we investigate the plasmonic effect from a quantum mechanical point of view, to more deeply understand the way behaviour of nanoantennas in terms of nanoscale radiations.

\subsection{Surface Plasmon Polaritons}
\subsubsection{Definitions and Wave Equation}
A fascinating phenomenon occurs when a conductor, such as a metal, comes into proximity with an insulator, known as a dielectric. Confined electromagnetic waves with a two-dimensional nature become capable of propagating along the interface. These fascinating waves are referred to as Surface Plasmon Polaritons (SPPs). The generation of SPPs arises from the intricate coupling between the electromagnetic fields and the oscillations of electron charges within the conductor.
SPPs have fascinating applications in nanophotonics \cite{SPP _ app 1}, optoelectronics \cite{SPP _ app 2}, and sensing \cite{SPP _ app 3} due to their unique ability to confine electromagnetic waves (light) at the nanoscale. In the absence of external sources, the nature of these waves can be explored through their wave equation \cite{Plasmonic Book 1}, i.e.,\\
\begin{align}\label{SPP Wave Equation}
  \nabla^{2}E-\frac{\varepsilon}{c^{2}}\frac{\partial^{2}E}{\partial t^{2}}=0,
\end{align}
where $c$ is the speed of light in vacuum.\\
Taking equation (\ref{SPP Wave Equation}) into the fourier domain, and considering $k_{0}=\omega / c$ as the wave vector, we obtain the \textit{Helmholtz Wave Equation}, i.e.,\\
\begin{align}\label{Helmholtz}
  (\nabla^{2}+\varepsilon k_{0}^{2})E=0 .
\end{align}
Applying the appropriate boundary conditions can offer solutions to (\ref{Helmholtz}) within various structures, which allows for two distinct sets of self-consistent solutions: TE modes (s-polarized) where the electric field aligns parallel to the interface, and TM modes (p-polarized) where the magnetic field aligns parallel to the interface. As an illustration, let us examine the most basic geometry suitable for supporting surface plasmon polaritons: a planar interface between a metal and a non-absorbent half space. We now derive the dispersion equation governing the propagation of waves within this arrangement. When considering the confined propagation of SPPs along the interface of two mediums, namely a conductor and an insulator, it is crucial to note that there must be a normal component of the electric field with respect to the surface. Consequently, the existence of s-polarized surface oscillations is negated, and the analysis should solely focus on TM modes. By considering propagation along the $x$ direction with a constant propagation constant $\beta$, and assuming homogeneity by setting $\frac{\partial}{\partial y}=0$, we can apply Maxwell's curl equations to derive the components of propagation. Through this analysis, we find that $E_{x}$, $E_{z}$, and $H_{y}$ are the only three components that are non-zero \cite{Plasmonic Book 1}, i.e.,
\begin{align}\label{TM mode_ Excitation}
  & E_{x,i}(z)= (-1)^{i}j A_{i}\frac{1}{\omega \varepsilon_{0} \varepsilon_{i}}k_{i}e^{j\beta x}e^{-k_{i}z}, \\
  & E_{z,i}(z)=-A_{1}\frac{\beta}{\omega \varepsilon_{0} \varepsilon_{i}}e^{j\beta x}e^{(-1)^{i+1}k_{i}z}, \\
  & H_{y,i}(z)=A_{i}e^{j\beta x}e^{(-1)^{i+1}k_{i}z},
\end{align}
where $\varepsilon_{2}$ represents a constant value, $\varepsilon_{1}$ is frequency-dependent, and $i = \{1,2\}$ depending on the region. It is important to note that in order to fulfill the metallic behavior of the environment, the real part of $\varepsilon_{1}$ must be negative ($Re\{\varepsilon_{1}\} < 0$). It is essential to acknowledge that in optical frequencies, the short wavelength of waves enables them to penetrate deeply into the metal, causing it to exhibit dielectric-like behavior. However, our intention is to address this phenomenon and prevent it from occurring. By ensuring that the waves gradually fade and following the condition $Re\{\varepsilon_{1}\} < 0$, we can achieve this objective. Lastly, in the regime of large wave vectors, the angular frequency and wave-vector of SPPs can be found in (\ref{Angular Frequency of SPP}), and (\ref{Wave vector of SPP}), respectively, i.e.,\\
\begin{align}
      & \omega_{spp}=\frac{\omega_{p}}{\sqrt{1+\varepsilon_{2}}} \label{Angular Frequency of SPP} , \\
      &k_{spp}=\frac{\omega_{spp}}{c}\sqrt{\frac{\varepsilon_{1}\varepsilon_{2}}{\varepsilon_{1}}+\varepsilon_{1}} , \label{Wave vector of SPP}
\end{align}
where $\omega_{p}$ denotes the \textit{plasma frequency}, as the threshold frequency for the dielectric behavior of metal.\\
Taking the surface plasmon polaritons, and consequently, the localized surface plasmons into the nanoantenna domain, one can comprehend their urgency for describing the behaviour of nan-scale antennas. SPPs contribute to the radiation enhancement in nanoantenna through their ability to concentrate electromagnetic energy, enhance light-matter interactions, facilitate directional radiation, achieve sub-wavelength resolution, and enable plasmon resonance tuning \cite{SPP_first,SPP_second,SPP_third}. Additionally, Localized Surface Plasmons (LSPs) contribute to radiation enhancement of nanoantennas through resonant absorption and scattering, field enhancement, the generation of hotspots, strong coupling with nearby emitters, and the ability to tune their resonance frequencies \cite{SPP _ ref 1,SPP _ ref 2,SPP _ ref 3,SPP _ ref 4,SPP _ ref 5,SPP _ ref 7}. Hence, SPP and LSP play crucial roles in designing nanoantennas for various applications, including sensing, imaging, and light manipulation at the nanoscale, and the physical explanation of their existence can be managed through quantum mechanics.


 Some of the most important ways to excite SPPs include through nanoantennas \cite{Nanoantenna _ Main Book,Nano-Optics Paper,Nanoantenna _ Main Review Paper,Nanoantenna _ Book 2}, prism-coupling \cite{SPP _ excitation 1,Plasmonic Book 1}, grating-coupling \cite{SPP _ excitation 2}, attenuated total reflection (ATR) \cite{SPP _ excitation 3}, and nanostructure-enhanced excitation \cite{SPP _ excitation 4}.

 Additional excitation techniques, and other structures supporting SPPs generation have been mentioned in the literature such as, Kretschmann Configuration \cite{SPP _ excitation 5}, Photonic Crystal-Based Excitation \cite{SPP _ excitation 6}, Two-Photon Excitation \cite{SPP _ excitation 7}, Nanostructured Metasurfaces \cite{SPP _ excitation 8}, which mostly share same fundamentals or overlap in some certain ways, while focusing on a specific application. For example, both the Kretschmann configuration and prism-coupling techniques involve using a prism to couple incident light into a metal-dielectric interface in favor of SPPs generation, and can sometimes be used interchangeably. Generally, each excitation method offers unique advantages and can be tailored for specific applications, ranging from sensing and imaging to light-matter interactions, and energy harvesting. Since the field of plasmonics is being explored and getting richer day by day, further advancements and novel applications are expected for SPPs in the near future.
\section{Nanoantennas and Terahertz Radiation} \label{THREE}
In this section, we explore the fascinating world of nanoantenna theory and search through the fundamental concepts of antenna parameters as we uncover the mechanisms behind these nano-scale structures with profound applications in various domains.
\subsection{Nano-antenna Theory}
\begin{figure}[!htbp]
  \centering
  \includegraphics[width=8.5cm, height=7cm,center]{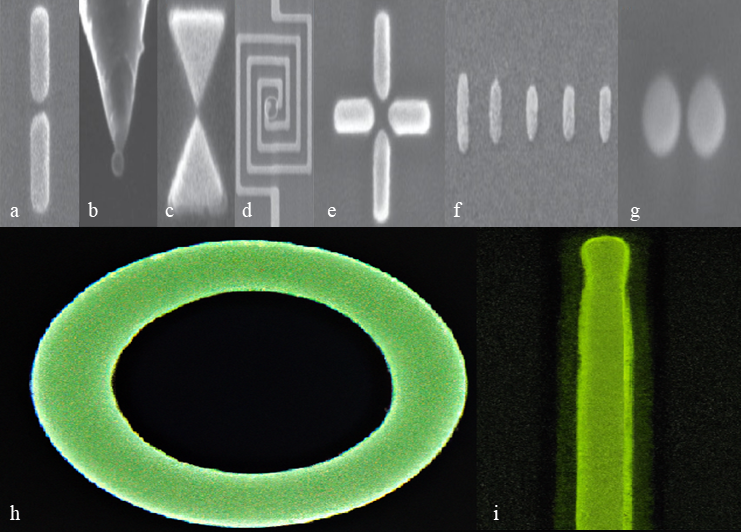}
  \caption{Scanning Electron Microscopy (SEM) images showcasing different types of optical antennas: (a) coupled-dipole antenna, (b) nanoparticle antenna, (c) bow-tie antenna, (d) square-spiral antenna, (e) cross antenna, (f) Yagi-Uda antenna, (g) Hertzian dimer antenna, (h) Nanoloop antenna, (i) Nanowire antenna. Images (a-g) captured from \cite{Nanoantenna _ Main Book}.}
  \label{figure : Nanoantenna Types}
\end{figure}

The emergence of 5G wireless communications, accompanied by research advancements over the past fifteen years \cite{5G first work 1,5G first work 2}, has assisted the development of meticulously crafted configurations. These configurations are tailored to provide the exponentially faster communication rates of the fifth-generation (5G) and forthcoming sixth-generation (6G) technologies. To enable their applications in various fields like biosensors, medical devices, energy harvesting, light manipulation, optical communications, and nanoscale communications, these configurations need the integration of miniaturized components at the nanoscale \cite{Nanoantenna _ Book 2}. In order to facilitate these advancements, it is imperative to design practical broadband antennas for both transmitters and receivers that can effectively handle these types of communications. Specifically, these modern antennas need to exhibit compact sizes, low profiles, high gains, wide bandwidths, and desirable radiation patterns. To keep pace with the rapid technological progress, nanoantennas are gaining prominence as they are primarily designed to operate within the terahertz range, enabling communication rates on the order of terabits per second. The nanoantenna serves as a crucial component responsible for the collection and absorption of electromagnetic waves with wavelengths that are proportional to its physical dimensions. By precisely tuning the size and shape of the nanoantenna, it becomes capable of effectively capturing and interacting with electromagnetic waves of specific wavelengths \cite{Nanoantenna _ Review paper 3}. Typically, a nanoantenna comprises a ground plane, a resonant cavity, and the transmitter/receiver section, which is the antenna itself. When electromagnetic waves with a specific frequency encounter the metal surface, they initiate the generation of Surface Plasmons (SPs) at the same frequency as the incident waves. The generated Alternating Current (AC) must be converted into a Direct Current (DC) to power an external load, as suggested by the transmission-line model. In other words, the absorbed incident waves are subsequently reflected and concentrated within the cavity using the ground plane section \cite{Nanoantenna _ Review paper 2,Nanoantenna _ Review paper 3}.
 As previously mentioned, the electromagnetic fields \textbf{E} and \textbf{H} are not subject to any additional conditions beyond the satisfaction of Maxwell's equations. This implies that as long as these equations are fulfilled, the propagation of electromagnetic waves is permitted, irrespective of their frequency. Nevertheless, classical Maxwell's equations are rendered inadequate at higher frequencies, such as in the optical range. This is attributed to the frequency variability of characteristic parameters in the participating media. Consequently, electromagnetic waves can penetrate the metal, inducing plasmon modes with wavelengths shorter than the free-space wavelength ($\lambda_{0}$), significantly changing the antenna's properties. \cite{Nanoantenna _ Main Book}.\\
Another difference between the classical and optical antenna theories lies in the feeding technology of the antennas. In radio wave antennas, impedance-matched transmission lines and waveguides, such as the coaxial cables, are being employed to feed these structures \cite{Classical Antenna _ Feed 1}. Within the optical regime,
the small dimensions of optical nanostructures make wiring between the antenna and the feed port (or transmitter/receiver) challenging. In this scenario, the transmitter or receiver can take the form of molecules, quantum dots, or tunnel junctions, connecting to the antenna through mechanisms involving energy or charge transfer \cite{Nanoantenna _ Main Book,Nanoantenna _ Book 2}.
The most important optical antennas that have been investigated in the literature as shown in figure (\ref{figure : Nanoantenna Types}) are metallic nanowires and nanoloops \cite{Nanowires _ Paper 1,Nanowires _ Paper 2,Nanowires _ Paper 3,Nanowires _ Paper 4,Nanowires _ Paper 5,Nanowires _ Paper 6,Nanowires _ Paper 7,nanoloop antenna _ paper 1,nanoloop antenna _ paper 2}, which form the main building blocks of other nanoantennas as well, coupled-dipole antennas \cite{coupled-dipole antenna _ paper 1,coupled-dipole antenna _ paper 2,coupled-dipole antenna _ paper 3,coupled-dipole antenna _ paper 4}, bow-tie antennas \cite{bowtie nanoantenna _ paper 1,bowtie nanoantenna _ paper 2,bowtie nanoantenna _ paper 3,bowtie nanoantenna _ paper 4,bowtie nanoantenna _ paper 5,bowtie nanoantenna _ paper 6}, hertzian dimer antennas \cite{hertzian dime nanoantenna _ paper 1}, nanoparticle antennas \cite{nanoparticle antenna _ paper 1,nanoparticle antenna _ paper 2,nanoparticle antenna _ paper 3,nanoparticle antenna _ paper 4}, yagi-uda nanoantennas \cite{yagi-uda nanoantenna _ paper 1,yagi-uda nanoantenna _ paper 2,yagi-uda nanoantenna _ paper 3,yagi-uda nanoantenna _ paper 4,yagi-uda nanoantenna _ paper 5,yagi-uda nanoantenna _ paper 6,yagi-uda nanoantenna _ paper 7}, cross antennas \cite{cross nanoantenna _ paper 1,cross nanoantenna _ paper 2}, and square-spiral nanoantennas \cite{square spiral nanoantenna _ paper 1}.
\noindent
\newline
\begin{table*}[ht]
\centering
\caption{Table of existing nanoantennas for different applications}
\label{tab:nanoantenna}
\resizebox{\textwidth}{!}{\begin{tabular}{ccccccc}
\toprule
\textbf{Structure} & \textbf{Material} & \textbf{Max Size (nm)} & \textbf{Central Wavelength (nm)} & \textbf{Gain (dB)/ Emission Intensity ($\mathbf{10^{6}}$ counts/s)} & \textbf{Bandwidth (nm)} & \textbf{Reference}  \\
\midrule
Nanowire & InP & diameter = 100 , length = 3130 & 850 & N.m & 10 & \cite{Nanowires _ Paper 1}\\
Nanowire & Gold/Si$O_{2}$ & N.m & 40-1630 & 911 &  25 & \cite{Nanowires _ Paper 2}\\
Nanowire & GaAs/InAs & diameter = 220 , length = 2000 & 950 & N.m & 70 & \cite{Nanowires _ Paper 3}\\
Nanowire & Au/GaAs & diameter = 220 , length = 2000 & 830 & N.m & N.m & \cite{Nanowires _ Paper 4}\\
Nanowire & Gold/ITO & 80 nm and 140 nm arms& 1100 & N.m & around 50 & \cite{Nanowires _ Paper 7}\\
Nanoloop & Silver & 500 & 1340 & 8.2 & 10 & \cite{nanoloop antenna _ paper 1}\\
Nanoloop & Gold & 500 & N.m & 7.5 & N.m & \cite{nanoloop antenna _ paper 2}\\
Bow-tie & Gold/Si$O_{2}$/ITO/PMMA & 500 & 820 & N.m & 100 & \cite{bowtie nanoantenna _ paper 1}\\
Bow-tie & Gold/Si/Cr & 200 & 1100(c$m^{-1}$) & variable & variable & \cite{bowtie nanoantenna _ paper 2}\\
Bow-tie & Gold/ITO/Glass & 300 & 780 & N.m/0.1 & 50 & \cite{bowtie nanoantenna _ paper 3}\\
Bow-tie & Gold/ITO/Glass & 575 & 660-808 & N.m & N.m & \cite{bowtie nanoantenna _ paper 4}\\
Bow-tie & SiC/Glass/GaN/Gold & 600 & 900(c$m^{-1}$) & N.m & 20(c$m^{-1}$) & \cite{bowtie nanoantenna _ paper 6}\\
Nano-particle & Gold & 60,100 & 740-1170 & N.m & ~10 & \cite{nanoparticle antenna _ paper 1}\\
Nano-particle & Gold & 20,40,80 & 633 & E.F=40 & 15,23 & \cite{nanoparticle antenna _ paper 2}\\
Nano-particle & Gold/Silver & 10 & 680 & N.m & 20 & \cite{nanoparticle antenna _ paper 3}\\
Yagi-Uda & Various(e.g. Si) & 150 & 520 & 12 & ~50 & \cite{yagi-uda nanoantenna _ paper 1}\\
Yagi-Uda & Gold/PC403 & 300 & 1500 & 20(Array),6(Single) & 100 & \cite{yagi-uda nanoantenna _ paper 2}\\
Yagi-Uda & Silver/Silica & 150 & 620 & 3 & 25 & \cite{yagi-uda nanoantenna _ paper 3}\\
Yagi-Uda & Si & 130 & 490,570 & ~8(at $\lambda=500 nm$) & 20 & \cite{yagi-uda nanoantenna _ paper 4}\\
Yagi-Uda & Gold/Glass/PMMA & 198(feed)/1200(substrate) & 1000 & ~9 & ~50 & \cite{yagi-uda nanoantenna _ paper 5}\\
Yagi-Uda & Silver/a-Si & 390(feed) & 1000 & N.m & 10 & \cite{yagi-uda nanoantenna _ paper 6}\\
Yagi-Uda & Gold/Ti$O_{2}$ & 40(feed) & 780 & ~7 & 50 & \cite{yagi-uda nanoantenna _ paper 7}\\
Cross & Si/quartz/Glass & 200 & 400-700 & N.m & 10 & \cite{cross nanoantenna _ paper 1}\\
Cross & Gold/Glass & 120 & ~800 & E.F=40 & 100 & \cite{cross nanoantenna _ paper 2}\\
Square-spiral & Gold/Ti/Ti$O_{x}$ & 5500 & 5-30 $\mu m$ & 2.5 & 10 & \cite{square spiral nanoantenna _ paper 1}\\
Coupled-dipole & Gold & 110 & 830 & N.m/1.15 & 50 & \cite{coupled-dipole antenna _ paper 1}\\
Hertzian-dipole & Gold/Silver/$Si_{3}N_{4}$ & 150 & 500 & 1.46 & ~30 & \cite{hertzian dime nanoantenna _ paper 1}\\
\bottomrule
\end{tabular}}\\
{\raggedright In this table, "N.m" indicates that the information is not mentioned in the specific reference. However, it is possible to deduce or estimate some values based on our background knowledge. For example, we might infer that a nanowire antenna provides a gain of around 2 dB. Additionally, in the sixth column, "E.F" represents the Efficiency Factor, measuring the nanoantenna's enhancement efficiency— a parameter assessing the structure's ability to enhance light-matter interactions.\par}
\end{table*}
Relying on these fundamental concepts, we may now define and look into the nanoantenna parameters in the next section.
\subsection{Antenna Parameters}
The evaluation metrics for optical nanoantennas closely are similar to those of classical antenna structures. However, due to the plasmonic effect and shorter wavelengths involved, they require recalibration. Specifically, this recalibration can be applied to any frequency band where quantum effects cannot be disregarded. Consequently, the following equations and definitions can be extended to other frequencies with the inclusion of specific considerations, if necessary.\\

\subsubsection{Radiation Pattern}
The radiation pattern provides a visual representation of an antenna's radiation properties in terms of its spatial coordination. Typically presented graphically, the pattern is defined within the spherical coordination system, i.e.,\\
\begin{align}
  &P_{rad}= \int_{0}^{\pi}\int_{0}^{2\pi} p(\theta,\phi)sin\theta d\phi d\theta ,
\end{align}
where $p(\theta,\phi)$ is the normalized power density. Once the necessary calculations have been performed, we can plot the radiated power as a function of $\theta$ or $\phi$ and analyze and report the resulting radiation pattern.\\
\subsubsection{Directivity}
Every antenna design aims to selectively transmit or receive propagations in specific directions while minimizing the influence of waves coming from other orientations. The directivity metric quantifies the antenna's ability to achieve this objective. It measures the antenna's performance in focusing transmitted or received waves. The directivity can be calculated within the spherical coordinate system using\\
\begin{align}
  & D(\theta,\phi)=\frac{4\pi}{P_{rad}}p(\theta,\phi) .
\end{align}
It is worth noting that the directivity can be defined independently for each axis, meaning that separate partial directivities can be calculated for the $\theta$-axis and the $\phi$-axis.\\
\subsubsection{Effciency}
Generally, the efficiency of an antenna is defined as\\
\begin{align}
   &\eta_{rad}=\frac{P_{rad}}{P_{rad}+P_{loss}} \label{antenna_efficiency} ,
\end{align}
in which the $P_{rad}$ , and $P_{loss}$ are the radiated power and the power dissipated to heat, respectively. The total power can be calculated utilizing the oscillating electric field \textbf{E} of the nanoparticle transmitter and the \textit{Poynting Vector, which determines both the value and the direction of power dissipation}. At the same time, the $P_{rad}$ is radiated from the entire nano-system, i.e., both the nanoantenna and the nanoparticle. 
\subsubsection{Gain}
\begin{figure*}[htp]
      \centering
      \subfloat[Single nano-particle]{
      \includegraphics[width=0.4\textwidth]{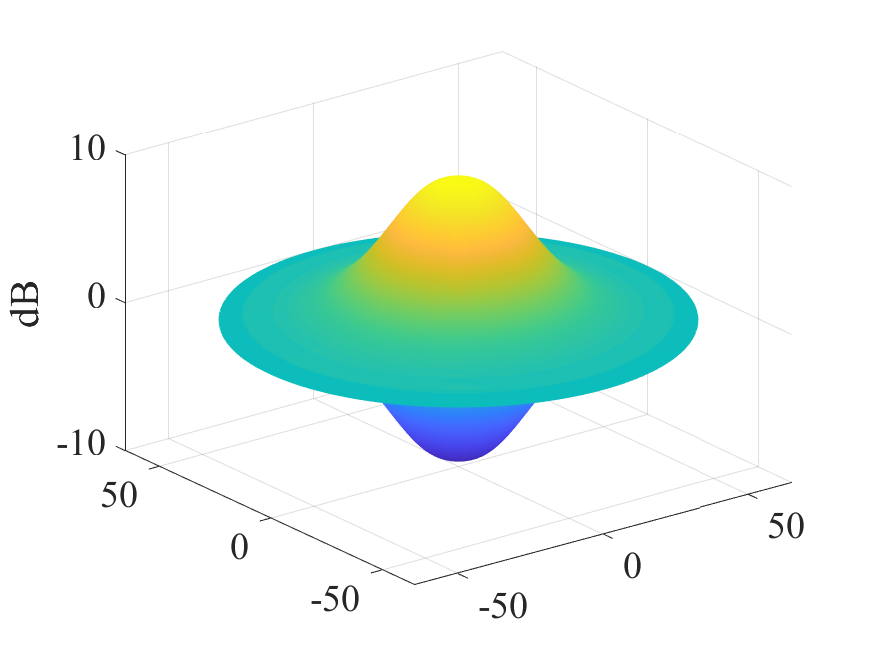}
      }
      \hspace{0.1\textwidth}
      \subfloat[4-elements array of nano-particles]{
      \includegraphics[width=0.4\textwidth]{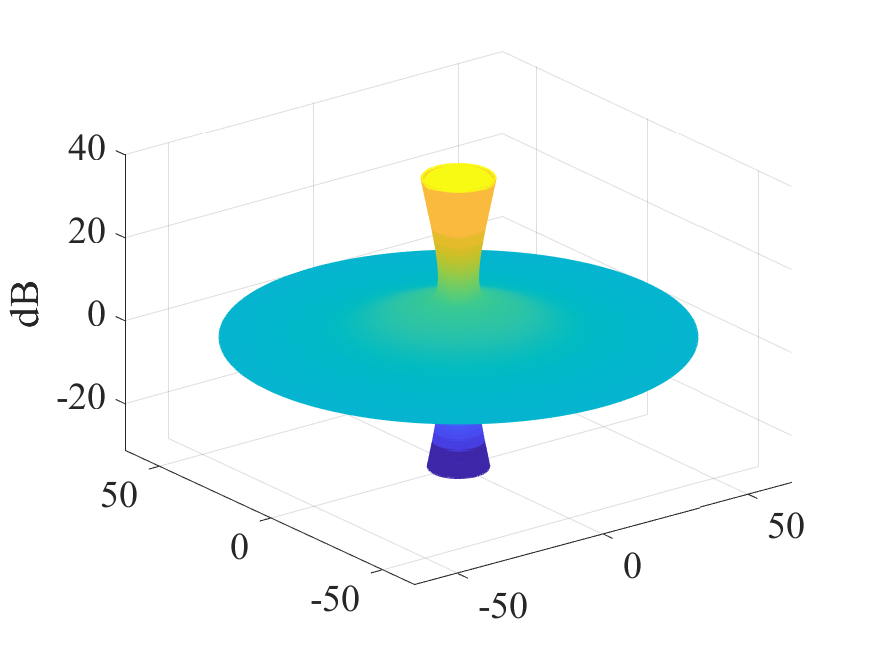}
      }
      \caption{Normalized Radiation Pattern : (a) A single nanoparticle functioning as a radiative nanodipole. As anticipated, this miniature component exhibits low gain and low directivity. (b) Four nanodipoles, when the maximum radiation in $\theta=0^{\circ}$ is desired ($\beta=-kd$). As expected, there is a noteworthy improvement in both directivity and gain, although the presence of null points is unavoidable.}
      \label{Dipole_pattern}
    \end{figure*}

The gain of an antenna can be defined as the ratio of the intensity in a specific direction (typically chosen as the direction of maximum radiation or maximum directivity) to the total input power dissipated by a hypothetical lossless isotropic reference antenna, whose gain is known, i.e.,
\begin{align}
  & G=\frac{4\pi}{P}p(\theta,\phi)=\eta_{rad}D ,
  \label{Directivity Equation}
\end{align}
where $p(\theta,\phi)$ is the normalized angular power density. A commonly used choice for the reference antenna is a dipole. For example, if we consider a nanoparticle with a current density represented by $J(r,t) \approx Re\{J(r)e^{-j\omega t}\}$ (which can be approximated as an oscillating dipole centered at the point-charge distribution of the nanoparticle) and a dipole moment denoted as $p(t) \approx Re\{pe^{-j\omega t}\}$, we can calculate the radiated power using \cite{Nanoantenna _ Main Book}\\
\begin{align}
  & P=\frac{|p|^{2}}{4\pi \varepsilon_{0}\varepsilon}\frac{n^{3}\omega^{4}}{3c^{3}} .
\end{align}
To calculate the normalized radiation pattern, we take into account that $n$ represents the dispersion-free index of refraction, and must be one for the sake of the causality of the system. Additionally, when considering the unit solid angle as $d\Omega = \sin\theta d\theta d\phi$, the normalized radiation pattern can be expressed as\\
\begin{align}
  & P_{n}=\frac{3}{8\pi}sin^{2}\theta ,
\end{align}
where $P_{n}$ denotes the normalized radiation pattern as visually shown in Fig. \ref{Dipole_pattern}. Consequently, we may calculate the directivity of this particular antenna using (\ref{Directivity Equation}).\\
\subsubsection{The Local Density of States}
One notable distinction between classical and optical antenna theory lies in the definition of \textit{input impedance}.
 In the context of optical nanoantennas, the concepts of current and voltage lack clear definitions, as the source is not directly connected to the antenna. Instead, since the power source is typically an emitter in the optical regime, we can calculate the \textit{Local Density of States} (LDOS) of the antenna. LDOS can be used to determine the density (or number) of states for the emitted photon out of the transmitter to occupy, allowing us to assess how its energy is dissipated \cite{Nanoantenna _ Main Review Paper}. As per this definition, a higher value of LDOS corresponds to better antenna performance. Thus, to enhance power dissipation, an optical nanoantenna must be optimized with higher values of LDOS. For instance, when considering a dipole emitter, we can calculate the LDOS using\\
\begin{align}
  & \rho_{p}(r_{0},\omega)=\frac{12\varepsilon_{0}}{\pi \omega^{2}}\frac{P}{|p|^{2}} ,
\end{align}
where $r_{0}$ and $P$ denote the location of dipole $p$, and the total radiated power, respectively \cite{Nanoantenna Impedance _ Paper}.\\
\subsubsection{Radiation Resistance}
The radiation Resistance is defined as\\
\begin{align}
  & R_{rad}=\frac{P_{rad}}{I_{max}^{2}/2} ,
\end{align}
where $R_{rad}$, $P_{rad}$, and $I_{max}$ are the radiation resistance, the radiation power, and the maximum current in the antenna's circuit, respectively.
As explained earlier, the antenna element can be represented as a load in the circuit model. In this context, the radiation performance of the antenna increases with a larger radiation resistance. Specifically, for a dipole antenna, the radiation resistance can be expressed as\\
\begin{align}
  & R_{rad}=\frac{2\pi}{3}Z_{0}( \frac{\Delta l}{\lambda})^{2} ,
\end{align}
where $\Delta l$ is the antenna length, $Z_{0} \approx 377 \Omega$ is the wave-impedance in free space \cite{Nanoantenna _ Main Book}.\\
\subsubsection{Antenna's Effective Wavelength}
When constructing an antenna, the design rules in both radio wave frequencies (including millimeter waves) and optical range illustrate a correlation between the operational frequency and the physical dimensions of the antenna. For instance, a half-wave antenna's length can be determined as $\Delta \ell = \frac{1}{2} \lambda$. In the case of array antennas, achieving maximum array efficiency requires careful spatial placement of the elements. This involves positioning the elements at distances proportional to the wavelength. A good example is the Hansen-Woodyard End-Fire Array, where the spacing parameter $d$ is given by $d=\frac{N-1}{N}\frac{\lambda}{4}$. Here, $N$ represents the number of array elements, and $d$ represents the required spacing between them \cite{Balanis Book}. In general, the essential dimensions of antennas can be expressed as $\ell$ = (constant value) $\times$ $\lambda$, emphasizing the linear relationship between antenna length ($\ell$) and the wavelength of radiation. However, in the case of optical frequencies, modifications need to be made as the previously mentioned equation no longer holds. This is because the PEC approximation is no longer valid at these frequencies.
Consequently, at optical frequencies, the antenna's response to incident waves changes due to the presence of a shorter effective wavelength $\lambda_{eff}$. This effective wavelength is determined by the material properties of the antenna, including the plasma frequency, conductivity, and penetration depth, i.e.,
\begin{align}
  & \lambda_{eff}=n_{1}+n_{2}\frac{\lambda}{\lambda_{p}}, \label{lambda_eff}
\end{align}
where $n_{1}$ and $n_{2}$ are constants depending on the antenna geometry, and $\lambda_{p}$ is the plasma wavelength \cite{effective lambda dipole 1}.\\
Calculating $\lambda_{eff}$ according to (\ref{lambda_eff}) can be a complex task, often requiring the use of numerical and experimental methods. However, the linearity of this equation indicates that classical antennas can theoretically be linearly miniaturized into optical antennas. This miniaturization can be achieved by utilizing a Scaling Ratio (SR), i.e.,
\begin{align}\label{scaling_ratio}
  & SR=\frac{\lambda_{eff} \lambda_{1}}{\lambda_{2}},
\end{align}
where $\lambda_{1}$ represents the operational frequency of the optical antenna, and $\lambda_{2}$ corresponds to the operational frequency of its classical counterpart \cite{Nanoantenna _ Main Book}.


In the following section, we inquire into the current applications of nanoantennas, while also introducing an emerging application, namely nano-scale radar systems.

\section{Applications of Nanoantennas} \label{FOUR}
  \subsection{6G Wireless Communications}
  Building upon the advancements of previous generations, such as 1G (commercialized in the 1980s) which provided limited voice calling capabilities and limited transfer rates, to the current development and commercialization of 5G in the 2020s, which offers applications for the Internet of Things (IoT) and massive broadband services with significantly higher rates of up to 10 Gbps, the forthcoming 6G is anticipated to assist in developing a fully-digital world with hyper data transfer rates of up to 1 Tbps. To highlight the significance of 6G, similar Key Performance Indicators (KPIs) or characteristics will be employed, as outlined in Table (\ref{Table _ KPIs}). When compared to the current most powerful existing generation, i.e., 5G, 6G offers noteworthy advancements. It is expected to be 100 times more reliable, possessing increased stability and dependability for various applications. Additionally, 6G offers data transfer rates that are 50 times faster, enabling even more efficient communication. Furthermore, 6G reduces latency by 10 times, resulting in significantly faster response times for in-time applications. These improvements in reliability, data transfer rate, and latency introduce 6G as a highly promising technology for future communication systems \cite{6gbook23_ref12,6gbook23_ref13,6gbook23_ref15}.
    \renewcommand{\arraystretch}{2}
  \begin{table}
     \centering
     \caption{Comparison between 6G and 5G KPIs \cite{6g _ book 23,6g _ book 10}}\label{6G and 5G KPIs}
   \begin{tabular}{|P{2.5cm}|P{2.5cm}|P{2.5cm}|}
   \hline
    \textbf{KPI} & \textbf{5G} & \textbf{6G} \\
     \hline
     \textit{Peak Data Rate} & $20$ $Gbps$ & over $1$ $Tbps$ \\
     \hline
     \textit{Experienced Data Rate}& $100$ $Mbps$ & $1$ $Gbps$ \\
     \hline
     \textit{Latency} & $1$ $ms$& $10-100$ $\mu s$  \\
     \hline
     \textit{Jitter} & Not Specified& lower than $1$ $\mu s$ \\
     \hline
     \textit{Enhanced Energy Efficiency} & Not Specified & $1$ $pJ/b$ \\
     \hline
     \textit{Reliability} & Error rate $< 10^{-5}$ & Error rate $< 10^{-7}$ \\
     \hline
     \textit{Enhanced Spectral Efficiency} & around $30$ $b/s/Hz$ & $100$ $b/s/Hz$ \\
     \hline
     \textit{Connection Density and Mobility} & $500$ $km/h$ & beyond $1000$ $km/h$ \\
     \hline
   \end{tabular}
   \label{Table _ KPIs}
  \end{table}
    To enable the advanced capabilities of 6G, it is essential to identify a suitable operational frequency range within the spectrum and establish a well-aligned infrastructure. The most promising and untapped frequency range for 6G implementation, which can support faster data transfer rates and wider bandwidth, is known as the \textit{Terahertz Gap}. This frequency range lies between 0.1 to 10 THz and offers exceptional potential for exceeding the limits of wireless communication in terms of speed and capacity. By exploring the capabilities of the Terahertz Gap, 6G can unlock new opportunities for hyper-connected and high-speed digital applications. The choice of the Terahertz Gap as the preferred frequency range for 6G implementation stems from several fundamental reasons. These reasons include limitations in the sub-6GHz band due to spectrum scarcity, insufficient bandwidth available in the millimeter wave (mmWave) range, constraints associated with the optical bands, and potential adverse effects of higher frequencies on the human body. These factors collectively highlight the need to explore alternative frequency ranges, such as the Terahertz Gap, to overcome these limitations and drive the development of 6G technology \cite{6G _ Survey one}. Terahertz waves possess several advantageous characteristics, making them well-suited for a range of applications. Their abilities include high resolution, the capability to penetrate non-conductive materials, sufficient bandwidth, and non-destructive properties—especially beneficial for medical purposes like cancer detection. These properties enable terahertz waves to provide important benefits and find applications in wireless communications and the upcoming 6G era.
    Along with many other usages, terahertz waves find applications in THz radars and sensing, enabling advanced navigation, collision avoidance for autonomous vehicles, and improved security screening \cite{THz _ cite 4}. They also offer wireless backhaul solutions, supporting the increasing demand for data transfer between base stations and core networks \cite{THz _ cite 5}.
     On the other hand, terahertz waves do face some limitations that should be taken into consideration. One of these limitations is the current lack of high-power THz wave transmitters. This means that the transmission distance of terahertz waves can be limited, affecting their range in certain applications. Additionally, terahertz waves are sensitive to high absorption coefficients caused by molecular absorptions in the propagation environment. This means that when terahertz waves interact with objects and obstacles, they can experience significant signal loss. This absorption phenomenon limits the ability of terahertz waves to penetrate certain materials and reduces their effectiveness in certain scenarios.
       To compensate for these limitations and harness the exceptional features of 6G communications, the implementation of highly directional antennas with sufficient gains and bandwidth is a must. These specialized antennas play a crucial role in overcoming the drawbacks associated with terahertz waves, enabling enhanced range and coverage. By offering high directivity, they focus the transmission beam in a specific direction, improving the efficiency of terahertz communication systems. Additionally, the large gains provided by these antennas boost signal strength, facilitating long-range communication capabilities. Furthermore, their wide bandwidth capabilities accommodate the high data rates required in 6G communications, supporting bandwidth-intensive applications and ensuring optimal performance. In this domain, antennas that operate in the frequency range of 0.1 to 10 THz are typically extremely small, often measuring in the nanoscale or sub-wavelength scale. Due to the reduced wavelength of electromagnetic waves within this range (ranging from 0.1 to 1 millimeter), conventional antennas designed for 4G or 5G communications (below 100 GHz) become ineffective.
         Hence, the field of antenna engineering and technologies faces numerous challenges in order to surpass these constraints and unlock the full potential of 6G applications, necessitating the utilization of nanoantennas.
   Terahertz antennas as the crucial components with the capability of addressing the challenges posed by 6G communications, can be divided into three major groups based on the manufacturing material: metallic or plasmonic antennas, novel antenna structures such as Metamaterial (MtM)-based and Graphene-based antennas, and dielectric antennas, which are outlined below, and compared in Table (\ref{Comparison_terahertz_antennas}).

   \begin{table*}[h]
\centering
\begin{tabular}{|c|c|c|c|c|c|}
\hline
\textbf{Antenna Type} & \textbf{Gain (dBi)} & \textbf{Compatibility} & \textbf{Average Size} & \textbf{Fabrication Difficulties} \\
\hline
\textit{Metallic Antennas} & High 20-30 & Limited & Sub-millimeter & Low \\
\hline
\textit{Dielectric Antennas} & Moderate 2-25  & Broad & Micrometer & Moderate \\
\hline
\textit{Graphene-based Antennas} & Extremely low $ < 1$ & Limited & Nanometer & Moderate \\
\hline
\textit{Metamaterial-based Antennas} & Moderate 2-5  & Limited & Sub-millimeter & High \\
\hline
\end{tabular}
\caption{Comparison Between Various Terahertz Antennas : \cite{thz_metallic 1,thz_metallic 2,thz_metallic 3,thz_metallic 4,thz_dielectric 1,thz_dielectric 2,thz_new materials 1,thz_new materials 2,thz_new materials 3,thz_new materials 4,thz_new materials 5,thz_new materials 6,thz_new materials 7,thz_new materials 8,thz_new materials 9,thz_new materials 10,thz_new materials 11,thz_new materials 12,thz_new materials 13,thz_new materials 14,thz_new materials 15,thz_new materials 16}}
Metallic antennas, Dielectric antennas, and New-material-based antennas (Graphene and Metamaterials). It highlights the key metrics such as lower cost and production complexity for metallic antennas, easy and advanced integration for dielectric antennas, and compatibility of new-material-based antennas with nano-channels.
\label{Comparison_terahertz_antennas}
\end{table*}

     \subsubsection{Metallic Antennae}
    Metallic antennas are manufactured using noble metals such as gold and silver, as well as copper, aluminium, or metal alloys that may include small percentages of palladium or platinum to improve stability and durability \cite{THz_Overview}. These antennas operate based on the principle of plasmonics, offering advantages such as low manufacturing cost, small sizes, and simple geometries. However, a major disadvantage of metallic antennas is their limited compatibility with planar structures, making it challenging to incorporate them into modern technologies. Furthermore, their use of mechanical adjustment methods for antenna positioning can reduce the overall performance \cite{THz_Overview}. An example of a metallic antenna is the Horn antenna, which may feature a tapered or conical design \cite{thz_metallic 1, thz_metallic 2}. Another instance is the travelling-wave corner cube antenna, which is characterized by simple open structures, low manufacturing requirements, and low coupling efficiencies \cite{thz_metallic 3, thz_metallic 4}. Despite their limitations, these types of antennas serve as practical options within the realm of metallic antenna technology.
     \subsubsection{Dielectric Antennae}
    Dielectric antennas possess several significant characteristics, including low input impedance, ease of fabrication and integration, and low cost \cite{THz_Overview}. Unlike metal antennas, dielectric antennas have minimal free-electron movement, resulting in reduced losses and improved propagation properties. Various geometries, such as butterfly, dual U-shaped, logarithmic periodic, log periodic sinusoidal, and bent-wire configurations, have been proposed using genetic algorithms \cite{THz_Overview, thz_dielectric 1}. However, one notable drawback of these antennas is the presence of surface wave effects. These effects occur when the radiation angle exceeds the cutoff angle, leading to trapped energy in the dielectric substrate and high energy loss. To overcome this issue, techniques such as loading lenses, reducing substrate thickness, and replacing the dielectric material with electromagnetic bandgap (EBG) structures have been employed \cite{thz_dielectric 2}.
    \begin{figure*}[htp]
    \centering
    \subfloat[Controlled drug-delivery from LTSLs by photothermal heating from MGNs in triple-negative breast cancer cells. The near infrared irradiations of the laser source with $\lambda = 808$nm is converted to heat by the multibranched gold nanoantennas (MGNs), and doxorubicin (DOX) is delivered from the LTSLs in breast cancer cells \cite{Medical 7}.]{%
        \includegraphics[width=0.45\textwidth]{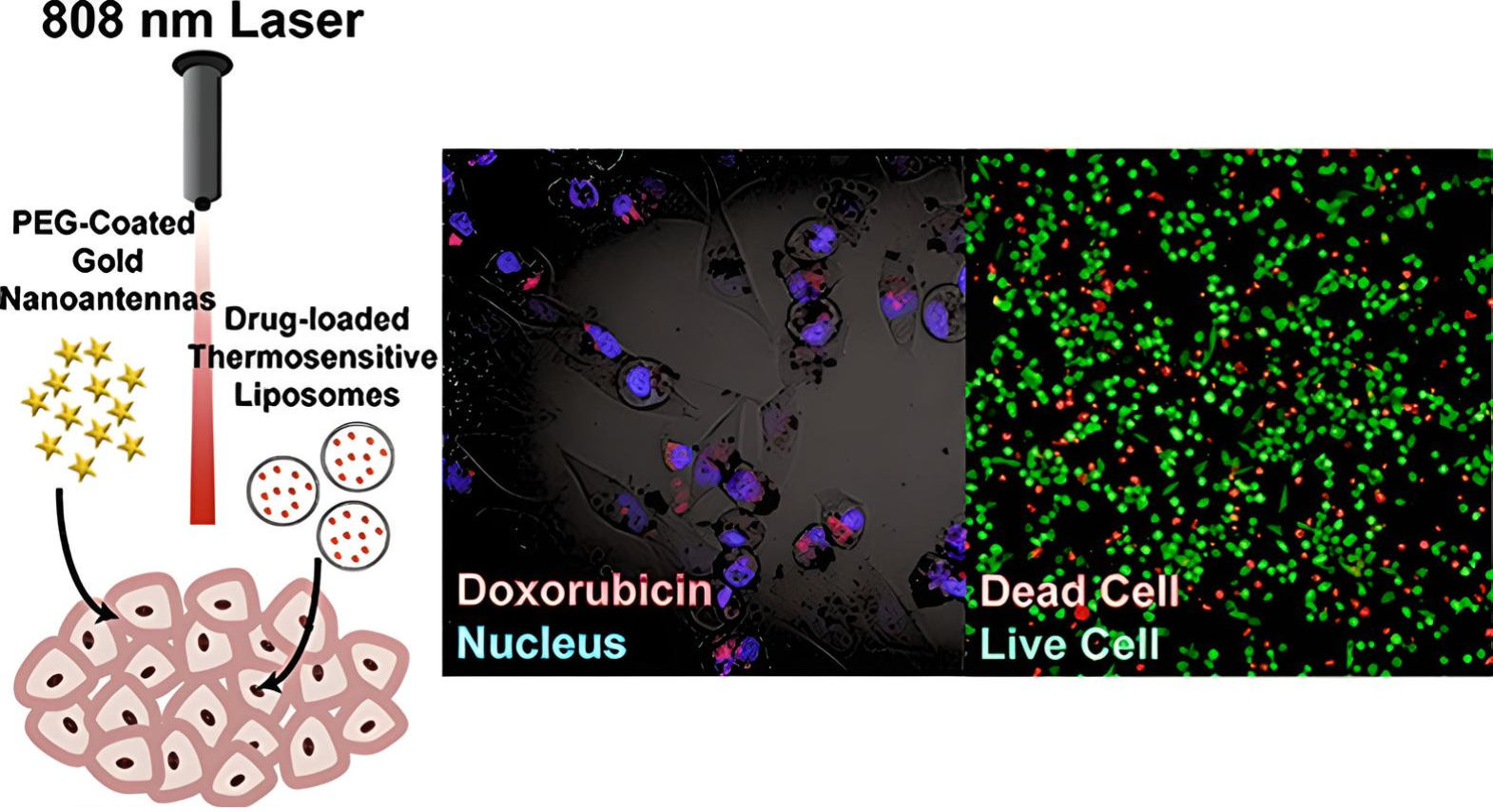}%
        }%
    \hspace{0.05\textwidth}%
    \subfloat[Ultrasensitive detection of Ebola virus (EBOV) antigens using a 3D plasmonic nanoantenna-based sensor as an on-chip immunoassay platform, for early-stage disease detection. The structure containing silica-based pillars between the gold nanodisks and backplane forms nanocavities, that can absorb laser irradiations at a certain wavelength. The figure on right, depicts the structure of EBOV sandwich assay on the nanoantenna array platform \cite{Medical 2}.]{%
        \includegraphics[width=0.45\textwidth]{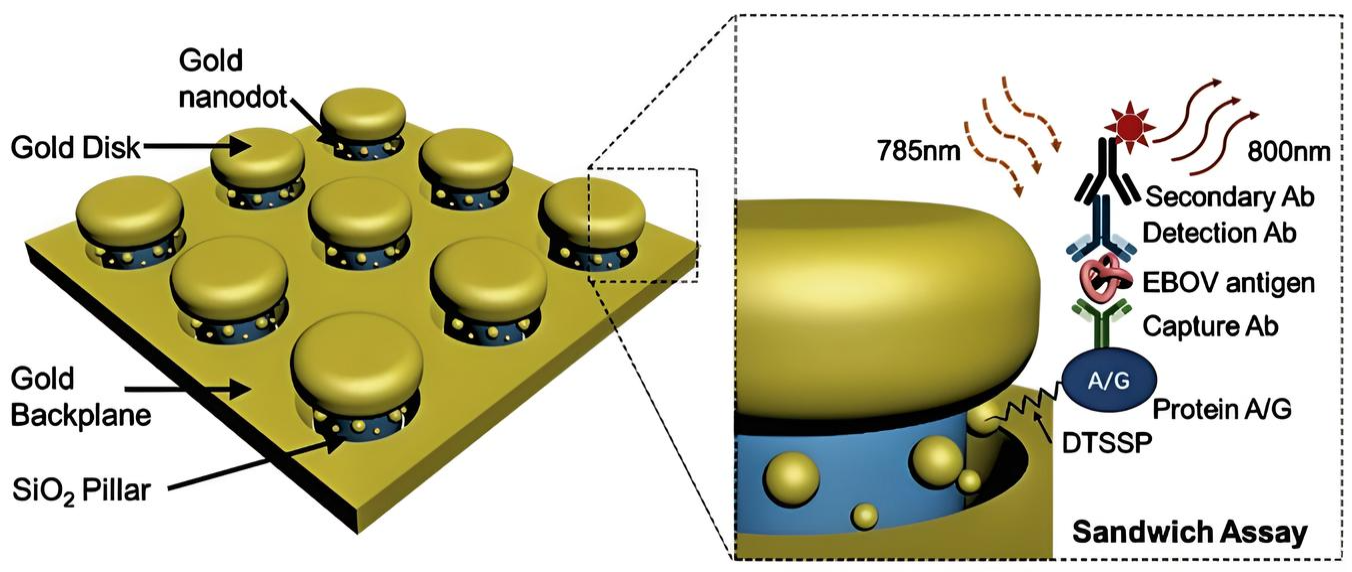}%
        }%
    \hfill
    \subfloat[Bull's eye structure with and without dot radiator, capable of forming a high-intensity focused beam and polarization-independent extraordinary optical transmission (EOT), applicable to biosensors. The figure on the bottom-left demonstrates the focused light beam from bull's eye structure without, and the middle-one illustrates the focused light beam with the existence of a dot radiator, respectively. The figure on the bottom-right shows the electric field amplitude at $\lambda = 800$nm, with and without the dot radiator \cite{Medical 4}.]{%
        \includegraphics[width=\textwidth]{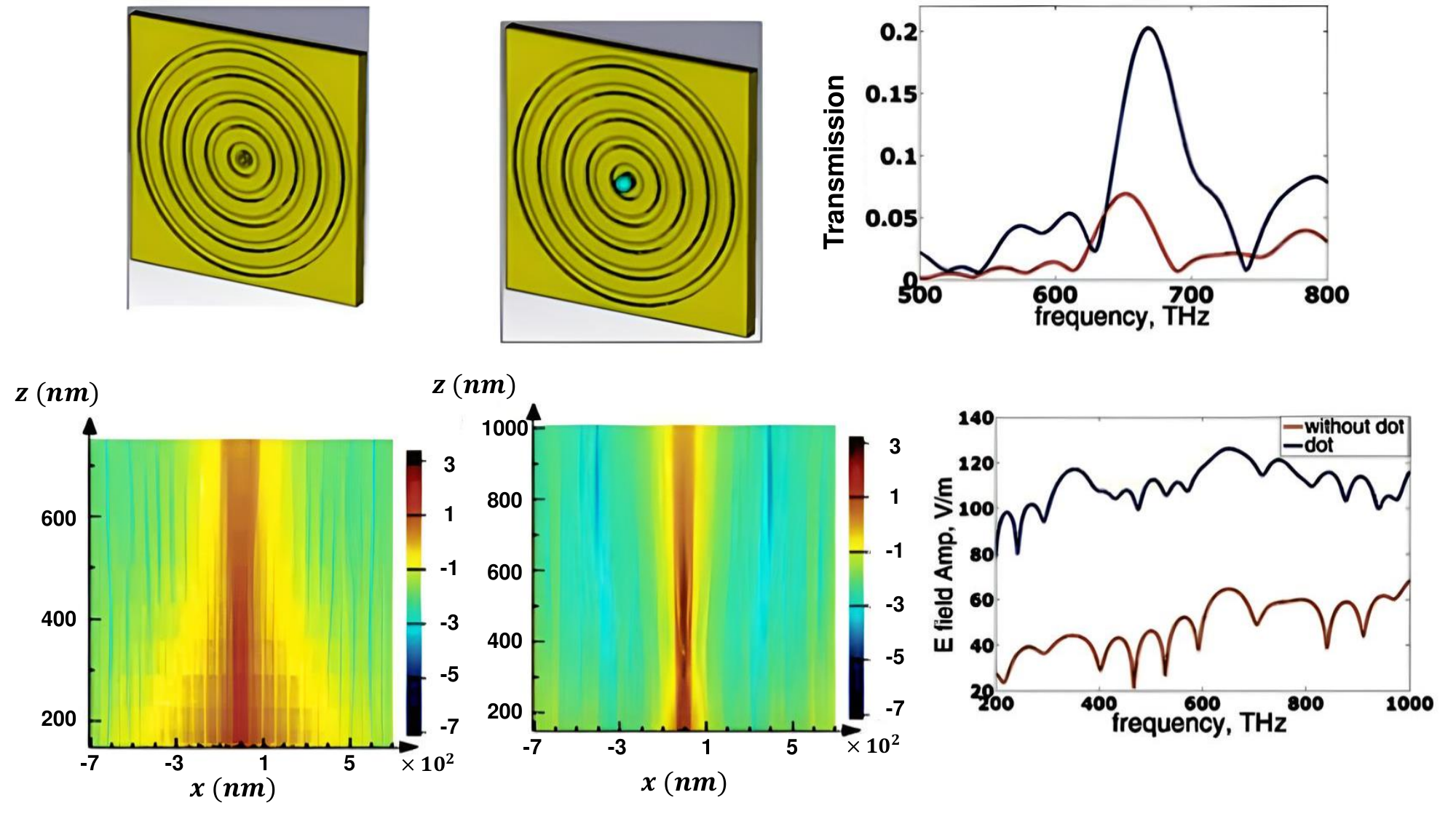}%
        }
           \caption{Different examples of using nanoantennas in medical applications}
    \label{Medical Applications}
\end{figure*}
     \subsubsection{Graphene-based and MtM-based Antennas}
    Currently, there are interesting developments in material science that propose using carbon nanotubes and metamaterials for new and improved dipoles. These innovations offer significant advancements in radiation performance compared to traditional options.
     For example, carbon nanotube dipoles exhibit resonances within a specific frequency range below the terahertz gap but rapidly lose effectiveness outside this range \cite{thz_new materials 1, thz_new materials 2}. Another promising option is bundles of carbon nanotubes wrapped in two dielectric layers, outperforming single-walled carbon nanotubes \cite{thz_new materials 3}. To further enhance bandwidth radiation characteristics and enable the fabrication of integrated antennas, researchers are exploring the application of two-dimensional planar materials, such as graphene, which possesses exceptional electrical properties. Graphene allows for dynamic control by adjusting bias voltage and facilitates the production of surface plasmonics. In contrast to graphene, metals cannot naturally couple with and convert electromagnetic waves in free space, leading to rapid attenuation of surface plasmon polaritons (SPPs) parallel to the metal-substrate interface \cite{thz_new materials 4}. Consequently, metallic antennas, which are subject to skin effects and have limited size, fail to meet the requirements of THz antennas. However, graphene exhibits a wide range of light absorption and regulation, emphasizing in-band transition dominance in the terahertz band. The collective oscillation of plasma in graphene enables excellent surface plasmon material properties, characterized by stronger binding and lower loss. Moreover, graphene allows for continuous electrical tuning \cite{thz_new materials 5, thz_new materials 7, thz_new materials 8}. Indeed, further research on new antenna materials will be crucial for achieving optimal performance. These novel materials can overcome traditional antennas' limitations, presenting advantages such as high gain, wide bandwidth, low loss, and cost-effectiveness. Therefore, alongside graphene, it is also worth exploring metamaterials and other two-dimensional materials in antenna design \cite{thz_new materials 15, thz_new materials 16}.

Typically, terahertz antennas discussed in the existing literature are on the scale of micrometers to millimeters. In other words, practical antennas that operate in the terahertz frequency range and are at the nano-scale are rarely investigated in current literature, which implies a high need for further research and development to manufacture nanoantennas suitable for terahertz communication. Most of the terahertz nanoantennas are constructed using graphene, but they face challenges such as low gains and efficiencies.
For instance, the use of bundled single-walled carbon nanotubes (SWCNTs) in the novel technology proposed in \cite{thz_new materials 2} aims to shift the operational frequency of the nano-dipole from higher values within the optical range to the terahertz gap. However, this approach exhibits a low performance in various radiative aspects, making it impractical for wireless communication components. Consequently, there is a critical need for advancements in the development of nanoantennas that can overcome these limitations and increase their feasibility for terahertz communication systems.
 \subsection{Nanoantennas in Medical Applications}
  Nanotechnology has opened the way for novel advancements in the field of medicine, and one of its most fascinating components is the utilization of nanoantennas. These tiny structures, designed to interact with light on the nanoscale, have emerged as promising tools for revolutionizing medical applications in different ways such as, imaging, biosensing and disease detection, drug delivery, photodynamic therapy (PDT), real-time monitoring, and photothermal therapy \cite{Medical 1,Medical 2,Medical 3,Medical 4,Medical 5,Medical 6,Medical 7,Medical 8,Medical 9,Medical 10}. For instance, nanoantennas are harnessed for innovative drug delivery strategies. Exploiting their plasmonic properties, they can generate localized heat when exposed to near-infrared light. As proposed by Yu-Chuan Ou et al. \cite{Medical 7}, this controlled photothermal effect triggers the release of therapeutic agents from liposomal drug carriers, like low-temperature-sensitive liposomes (LTSLs). By co-delivering multibranched gold nanoantennas (MGNs) and LTSLs, targeted drug delivery is achieved with minimal damage to healthy tissue. The MGNs' unique geometry enhances light-to-heat conversion efficiency, enabling controlled drug release. This approach offers several advantages: precise drug delivery at the tumor site, noninvasiveness, and improved therapeutic efficacy in aggressive conditions like triple-negative breast cancer. MGN-mediated photothermal hyperthermia overcomes multidrug resistance and enhances drug delivery. This combination of nanoantennas and LTSLs opens avenues for clinically relevant noninvasive drug delivery platforms with potential to revolutionize cancer treatment \cite{Medical 7}.
    Nanoantennas have emerged as a groundbreaking tool in molecular sensing and detection as well, as exemplified by Zang et al., in \cite{Medical 2}, when in the context of detecting the Ebola virus (EBOV) antigen, the significance of using nanoantennas is of greatest importance. The urgency to combat highly lethal pathogens like Ebola underscores the need for precise and sensitive diagnostics. Current methods, such as reverse transcriptase polymerase chain reaction (RT-PCR) and immunoassays, pose limitations in terms of sensitivity and controlled environments, and this is where nanoantennas shine. These plasmonic nanostructures possess the unique ability to interact intensely with biological elements due to their nanoscale characteristics. Through innovative techniques like nanoimprint lithography, Zang et al. have constructed 3D nanoantenna arrays. These structures exhibit optical resonance, magnifying fluorescence signal levels for early antigen detection. This leap in sensitivity is staggering, when detection of EBOV soluble glycoprotein (sGP) in human plama down to $220$ fg m$L^{-1}$ is achieved, with improvements of up to 240,000-fold compared to the $53$ ng m$L^{-1}$ EBOV antegien detection limit of the standard immunoassays. The scalable fabrication process further enhances their applicability, aligning with established assay formats. Zang's work showcases the transformative potential of nanoantennas, not only for Ebola but as a universal platform for diagnosing an array of diseases with unparalleled sensitivity and precision \cite{Medical 2}.

    Numerous other publications within the literature demonstrate various applications involving nanoantennas and nanostructures in the realm of medical science. One such instance emphasizes on the employement of dielectric dot radiators, represented by the optical dot antenna (ODA), as a flexible way for arranging electromagnetic properties, thereby supporting transmission and directivity within bull's eye structures applicable to biosensors and nanophotonics \cite{Medical 4}. Equally notable is the integration of plasmonic nanoantennas, manufactured by the indium-tin oxide (ITO) nanorod arrays, which, as an alternative to conventional plasmonic materials like gold and silver, avoid their inherent limitations—such as high loss and cost \cite{Medical 1}.
     These examples are only a few out of many in the literature. However, it remains evident that further dedicated research endeavors are required to develop the maturity of this field.
  \subsection{Nanoantennas for Enhanced Sensing in Nanoradar Systems}
      \begin{figure*}
      \centering
      \includegraphics[width=6.5in,height=3in]{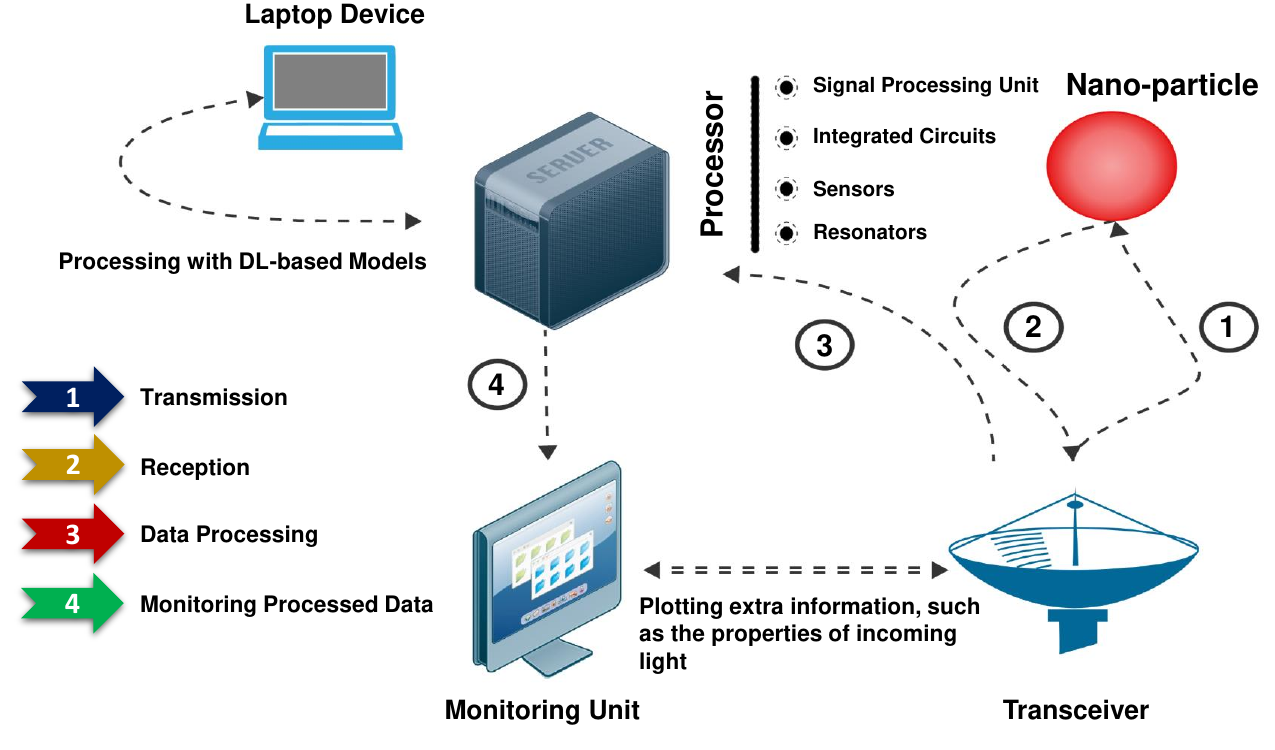}
      \caption{The NR system Block Diagram comprises seven main blocks: Antenna and Transceiver responsible for wave transmission and reception, Signal Processing Unit for processing and analysis of received signals, Monitoring for displaying and visualizing processed radar data and other information, Sensors for providing feedback and control signals to the monitoring unit, Integrated Circuits (ICs) for controlling actuator movements and receiving data from sensors, and Resonator for ensuring frequency stability. All of these components can be integrated into a unified processor unit.}\label{NsR diagram}
    \end{figure*}
  \subsubsection{Discussions}
        A radar system is an electromagnetic technology employed for the detection and tracking of objects, specifically those in motion. This technology involves transmitting electromagnetic waves and subsequently analyzing the signals that are reflected or backscattered by the objects. Through this process, a radar system can unveil various characteristics of the objects, including their distance, speed, direction of movement, as well as material and physical properties. This system typically employs essential components such as antennas or transceivers, along with processing units and monitoring units. These components collaborate with each other, to interpret the received signals from the targets, thereby enabling the determination of their respective locations, speeds, and other relevant features.
        At present, there is limited literature available on nanoradars, with only a few notable works being published \cite{Nanoradars _ 1,NsR example 2,NsR example 3,NsR example 4}. In addition to these limited works, there have been various discussions on sensing, imaging, and diverse forms of nano detections in the literature \cite{Lead to NsR 1,Lead to NsR 2,Lead to NsR 3,Lead to NsR 4,Lead to NsR 5}. These discussions provide valuable insights that can contribute to the development of a nanoradar system. Integrating these discussions with the existing works on nanoradars will help establish a more comprehensive understanding of the concept. In this section, we explore potential methodologies for nanoradar systems. Generally speaking, \textbf{A Nano-scale Radar (NR)} is a compact system designed to track minuscule entities, such as molecules or ions, at the nanoscale level, with the unique aspect of operating in optical frequencies, and to perform this task, they must be capable to detect and analyze the back-reflected/back-scattered waves from the targets efficiently. Unlike conventional radar systems, which explore the transmission, reflection, and reception of signals based on the Doppler Effect—a domain well-explored in literature—important challenges can occur at the nanoscale. Specifically, due to the reliance on the optical and quantum properties of targets for reception and detection, such as scattering properties of the nanoparticle, a designed nanoradar might exhibit reliable performance in detecting one specific type of molecule yet struggle to discern another, while these optical properties can vary noticeably from one material to another. In other words, an anticipated potential NR system must be well-designed and optimized for only one particular application, in one particular nano-channel. In order to study the scattering properties of nanoparticles different methods, such as Mie-Theory or Rayleigh-Gans-Debye-Theory are proposed \cite{Mie Theorem _ 1,Mie Theorem _ 2,RGD Theorem _ 1,RGD Theorem _ 2}. These theorems investigate the backscattered light (or electromagnetic waves) from the nano-objects, by taking into account their sizes (relative to the wavelength of operation), shapes, and refractive indices. The Mie theory is a comprehensive approach for calculating light scattering by particles of all sizes, involving complex calculations based on Bessel functions and spherical harmonics \cite{Mie Theorem _ 1}. It provides accurate results for various particle sizes and materials.
        Mathematically speaking, let's consider an abstract and simplified scenario, where there is only one single spherical scatterer existing in the environment, exposed by a dipole nano-radiator. In this case, we can find closed-mathematical expressions for the incident and scattered electric fields \cite{Mie Theorem _ 1,Mie Theorem _ 2}, i.e.,
    \begin{align}\label{Mie_incident}
    & E_{i}(\textbf{r},\omega)=\sum_{\ell = 0}^{\infty}\sum_{n = -\ell}^{\ell}(p_{\ell n}N_{\ell n}^{(1)}(k \textbf{r})+q_{\ell n}M_{\ell n}^{(1)}(k \textbf{r})),\\
    & E_{s}(\textbf{r},\omega)=\sum_{\ell = 0}^{\infty}\sum_{n = -\ell}^{\ell}(a_{\ell n}N_{\ell n}^{(3)}(k \textbf{r})+b_{\ell n}M_{\ell n}^{(3)}(k \textbf{r})),
    \end{align}
    \vspace{0.5 cm}
    where $N_{\ell n}^{(1)}$ and $M_{\ell n}^{1}$ are the vector spherical harmonic basis
    \newpage
    \noindent
     functions for electric (TM), and magnetic (TE) harmonics, respectively, and $a_{\ell n}$, $b_{\ell n}$, $p_{\ell n}$, and $q_{\ell n}$ are the coefficients, that must be calculated. Here, superscript $(1)$ means that the coefficients are captured from spherical Bessel functions. If the nano-particle is located at $\textbf{r}_{0}$, the scattering coefficients $a_{\ell n}$ and $b_{\ell n}$ can be calculated, i.e.,
    \begin{align}\label{Mie_coefficients_single}
      & a_{\ell n}=(-1)^{n}\frac{\textit{i}\alpha_{\ell} k^{3}}{\varepsilon}\frac{2\ell +1}{\ell (\ell +1)}N_{\ell (-n)}^{(3)}(k\textbf{r}_{0}).\textbf{p}, \\
      & b_{\ell n}=(-1)^{n}\frac{\textit{i}\beta_{\ell} k^{3}}{\varepsilon}\frac{2\ell +1}{\ell (\ell +1)}M_{\ell (-n)}^{(3)}(k\textbf{r}_{0}).\textbf{p},
    \end{align}
     where \textbf{p}, $\alpha_{\ell}$, $\beta_{\ell}$, $k$, and $\varepsilon$ are dipole moment, Lorentz-Mie single sphere coefficients, wavenumber, and the dielectric function, respectively. Here, superscript $(3)$ means that the coefficients are captured from spherical Hankel functions.  The computational cost, for calculating these coefficients can be dramatically high, which makes finding the scattering fields extremely difficult, specifically when a more extended scenario is studied with assuming multiple nano-particles in the environment, i.e.,
     \begin{align}\label{Mie_coefficients_multiple}
       & a_{\ell n}^{i}=\alpha_{\ell}^{i}\{\frac{\textit{i} k^{3}}{\varepsilon}N_{\ell (-n)}^{(3)}(k \textbf{r}_{0}^{i}). \textbf{p} \\ \nonumber
       &+ \sum_{\substack{j=1 \\ j \neq i}}^{N_{s}}\sum_{\ell'=1}^{\infty}\sum_{n' = \ell'}^{\ell'}[a_{\ell'n'}^{j}A_{\ell n\ell'n'}^{(3)}(k \textbf{R}^{ji})+a_{\ell'n'}^{j}B_{\ell n\ell'n'}^{(3)}(k\textbf{R}^{ji})]\},\\
       & b_{\ell n}^{i}=\beta_{\ell}^{i}\{\frac{\textit{i} k^{3}}{\varepsilon}M_{\ell (-n)}^{(3)}(k \textbf{r}_{0}^{i}). \textbf{p} \\ \nonumber
       &+ \sum_{\substack{j=1 \\ j \neq i}}^{N_{s}}\sum_{\ell'=1}^{\infty}\sum_{n' = \ell'}^{\ell'}[b_{\ell'n'}^{j}B_{\ell n\ell'n'}^{(3)}(k \textbf{R}^{ji})+b_{\ell'n'}^{j}B_{\ell n\ell'n'}^{(3)}(k\textbf{R}^{ji})]\},
     \end{align}
     where $A_{\ell n\ell'\acute{n}}^{(3)}$ and $B_{\ell n\ell'n'}^{(3)}$ are vector harmonic addition coefficients to model the couple between $i^{\text{th}}$ and $j^{\text{th}}$ spherical nano-particles, and $\textbf{R}^{ji}= \textbf{r}-\textbf{r}_{0}$. Note that in the case of a plane-wave excitation, $m = \pm \ell$, which is not valid for dipole excitation.
     In contrast to the Mie theory, the Rayleigh-Gans-Debye (RGD) theory simplifies scattering calculations for small particles (radius $\ll$ wavelength) by assuming isotropic polarizability and straightforward scattering patterns \cite{RGD Theorem _ 1}. The RGD approximation is applicable under specific conditions, i.e.,
     \begin{align}\label{RGD_Conditions}
       & |n-1| \ll 1, \\
       & kd|n-1| \ll 1,
     \end{align}
     where $d$ is the linear dimension of the particle, $n$ stands for the relative complex refractive index of the particle, with respect to the surrounding medium, and $k$ is the wavevector. Considering an arbitrary shaped particle exposed to a plane wave radiation in z direction, the parallel and perpendicular components of the scattered electric field can be calculated using the scattering matrix \cite{RGD Theorem _ 1,RGD Theorem _ 2}, i.e.,
     \begin{align}\label{RGD_equation}
           \begin{pmatrix}
             \Delta E_{\parallel scat} \\
             \Delta E_{\perp scat}
           \end{pmatrix} =\frac{exp(\textit{i}k(r-z))}{-\textit{i}kr}
               \begin{pmatrix}
                 S_{2} & 0 \\
                 0 & S_{1} \\
               \end{pmatrix}
                       \begin{pmatrix}
                         E_{\parallel inc} \\
                         E_{\perp inc} \\
                       \end{pmatrix},
     \end{align}
     where $E_{\parallel scat}$, $E_{\perp scat}$, $E_{\parallel inc}$, and $E_{\perp inc}$ are the parallel and perpendicular components of the scattering, and incident fields, respectively. Also, the scattering matrix elements needs to be calculated separately, i.e., $$S_{1}= -\frac{\textit{i}k^{3}}{2\pi}(n-1)Vf(\theta,\phi),$$ $$S_{2}=-\frac{\textit{i}k^{3}}{2\pi}(n-1)Vf(\theta,\phi)cos(\theta),$$ $V$ is the volume of the particle, and $f(\theta,\phi)$ is the form factor, i.e.,
     \begin{align}\label{Form_factor}
     & f(\theta,\phi)=\frac{1}{V}\int_{V}e^{\textit{i}\delta}d\nu,
     \end{align}
     where $\delta$ is the phase, and can be calculated for different cases, e.g., for a homogeneous sphere $\delta = 2k \xi sin (\frac{\theta}{2})$, where the variable $\xi$ is the distance from the origin to a plane of constant phase. Note that, in the case of a heterogeneous particle including $j$ homogeneous regions, (\ref{RGD_equation}) must be generalized by calculating $S_{1}$ and $S_{2}$ for each region, and summing up the results to achieve the final terms of the scattering matrix elements.\\
     RGD theory is suitable for quick estimations but lacks accuracy for larger particles, while Mie theory covers a broader range of particle sizes and materials with more accurate predictions. Table (\ref{Comparison Between Mie and RGD}) provides a brief comparison between these two.
     Due to the noticeable computational costs associated with both Mie and Rayleigh-Gans-Debye approaches, it becomes essential to use software tools such as Ansys Lumerical or COMSOL to fully simulate the entire procedure physically, or MATLAB to explore the patterns of back-scattered electromagnetic waves numerically.
     \begin{figure*}[h]
        \centering
        \subfloat[Scattering intensity by Mie theory]{
        \includegraphics[width=0.5\textwidth]{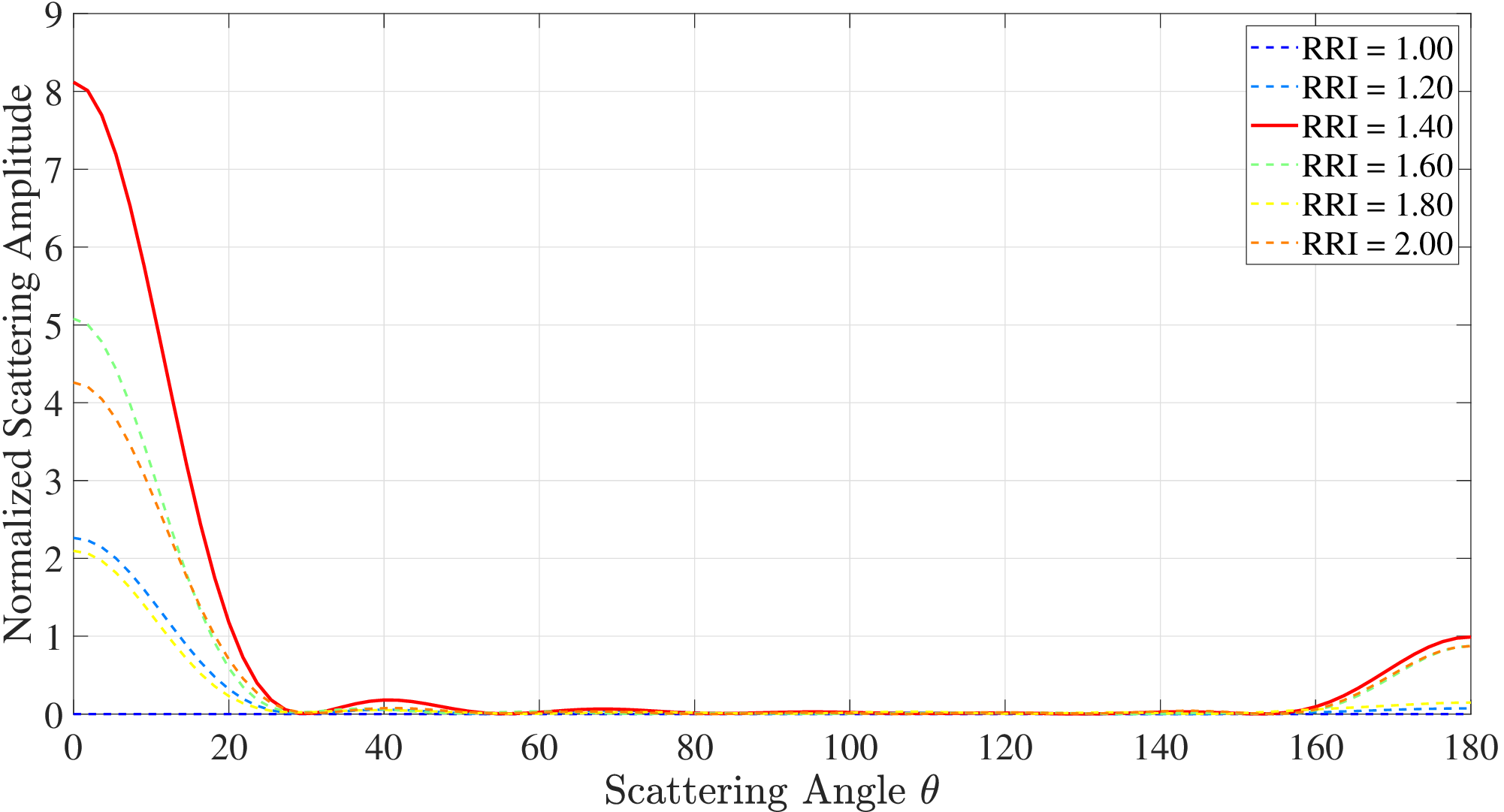}
        }
        \subfloat[Scattering intensity by RGD theory]{
        \includegraphics[width=0.5\textwidth]{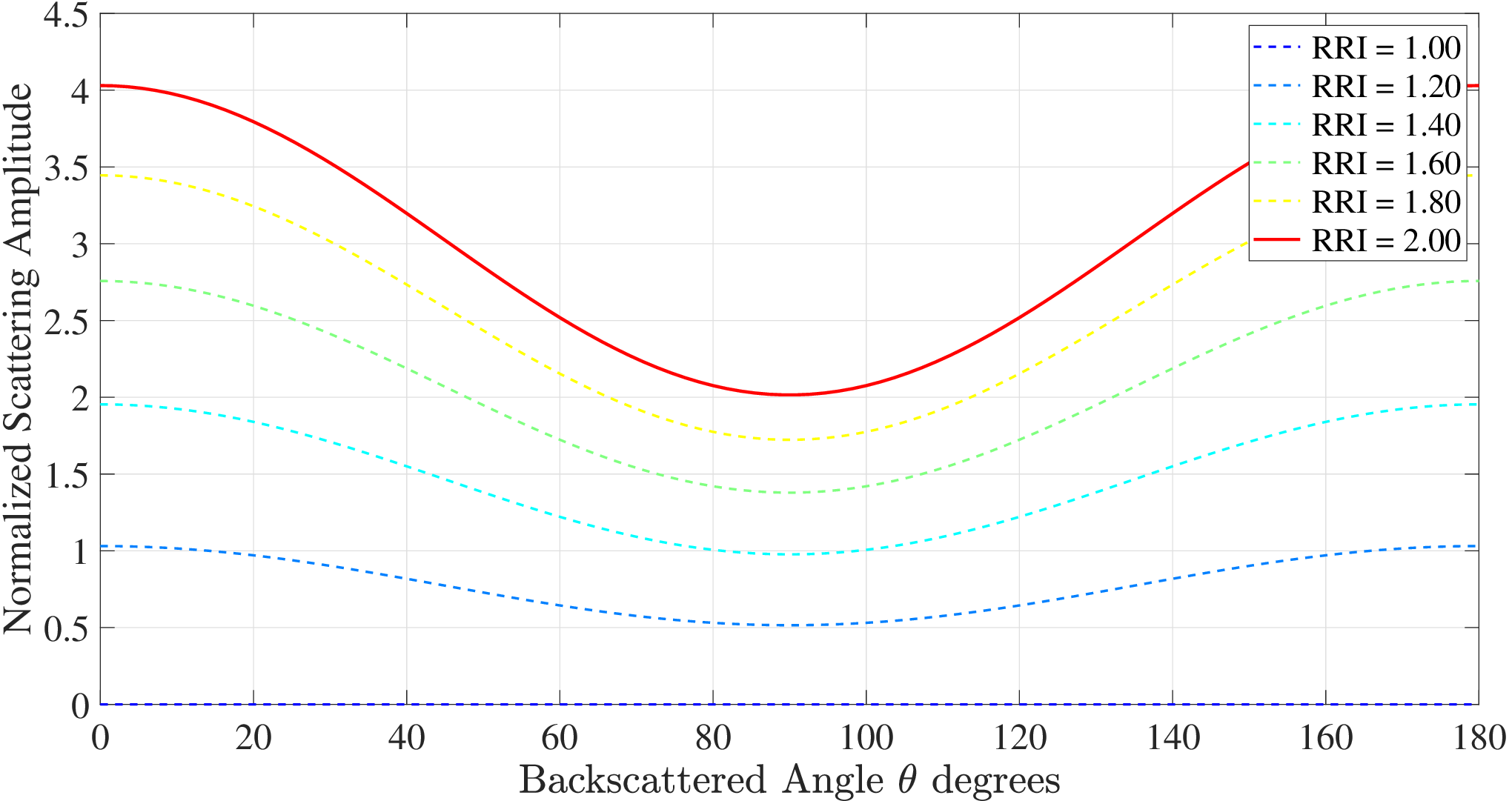}
        }
        \caption{(a) The intensity of the back-reflected electromagnetic waves as a function of the scattering angle, for a homogeneous particle having a radius approximately equivalent to the incident light's wavelength (around 500 nm and 428 nm, respectively). In this case, (\ref{Mie_coefficients_single}) can be used. (b) The intensity of the scattered waves as a function of the scattering angle, for a homogeneous spherical particle with a radius much smaller than the incident light's wavelength, i.e., r $\ll \lambda$ (around 50 nm, and 428 nm, respectively). In these figures, RRI is the abbreviation of Relative Refractive Index, and denotes the ratio of the refractive index of the nano-particle to the surrounding medium.}
        \label{Mie and RGD results 1}
    \end{figure*}
     \subsubsection{Simulations and Analysis}
\begin{figure}
      \centering
    \includegraphics[width=4in,height=2.5in]{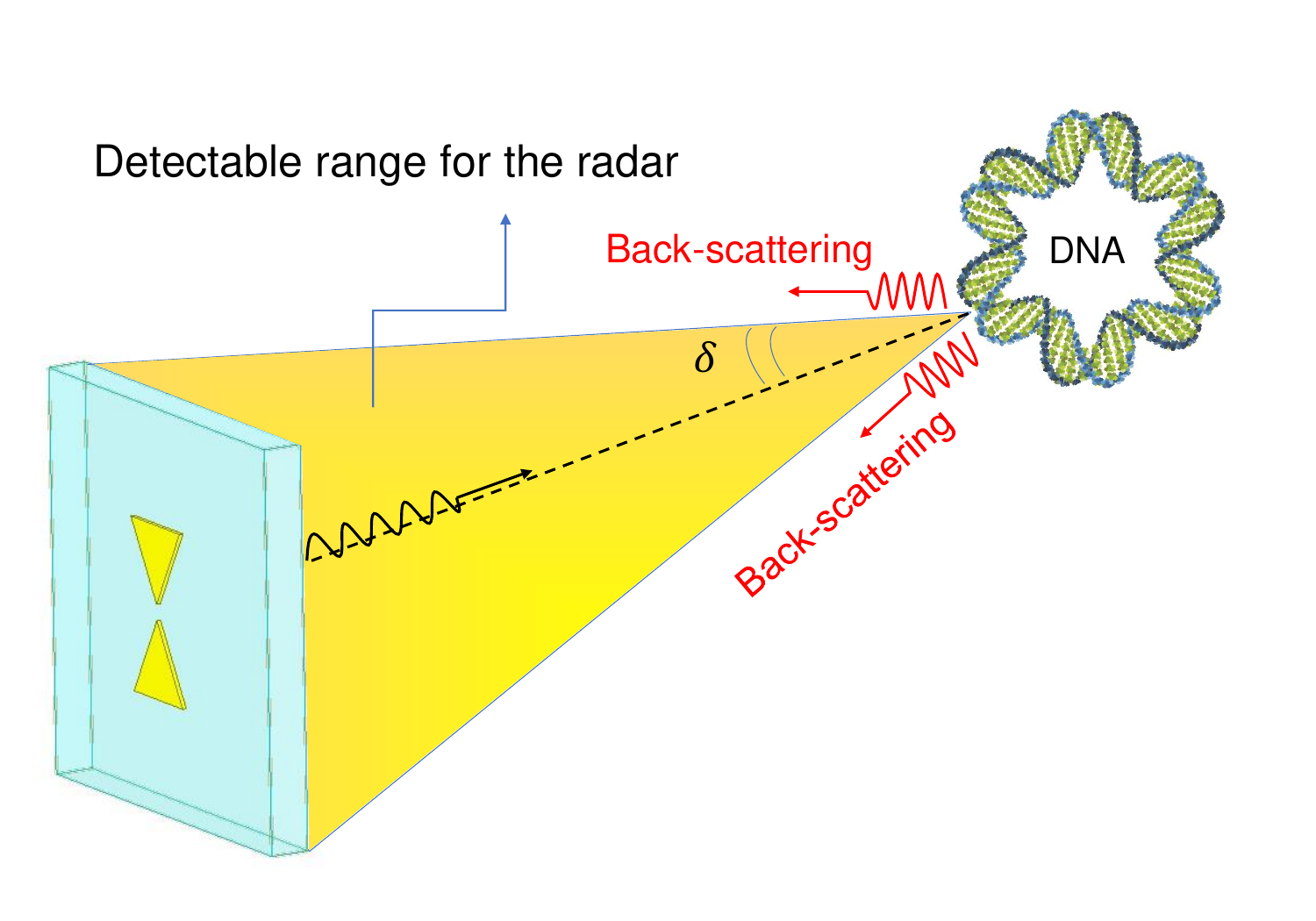}
    \caption{A hypothetical configuration of the Nano-scale Radar System containing a DNA molecule as the scatterer (target), and a bow-tie structure as the transmitting antenna is depicted, showcasing the detectable angle $\delta$. Due to the symmetry, the total feasible range of reflection angles for radar detection spans $2\delta$.}
    \label{NsR configuration}
    \end{figure}

      In this section, MATLAB has been used to find the scattering coefficients, and investigate the scattering patterns, and the results are depicted in Figure \ref{Mie and RGD results 1}.
      In this context, consider an optical nano dot functioning as a transmitting antenna, emitting at a frequency of $700$ THz (wavelength = 428 nm), within a nano-channel filled with Air ($\varepsilon=1$). The objective is to detect a single homogeneous spherical object under two distinct scenarios. In one scenario, the nanoparticle's radius is set at $50$ nm, i.e., the Mie theory must be applied, while in the other, is adjusted to $500$ nm, i.e., RGD theory is employed.
    The key difference between these two cases is the ratio of the target's size to the operational wavelength. In the first scenario, size $\ll$ wavelength is met, allowing for the utilization of the Rayleigh-Gans-Debye approximation. Conversely, in the second scenario, where this condition does not hold, the more sophisticated Mie's algorithm must be employed to accurately model the scattering phenomenon. As shown in Figure \ref{Mie and RGD results 1}, due to their minuscule dimensions, nano-particles exhibit limited capability in reflecting light with high intensities. Consequently, the received signals (\textit{echo-signals}) are weak, specially within certain scattering angle intervals, such as between 50 to 150 degrees, which makes the process of detecting and analyzing the back-reflected signals, practically complex for the processing unit of the nanoantenna. One way to overcome this issue, is to increase the refractive index of the particle, as its inherent physical aspect. This index signifies the ability of the nano-particle to repel incident waves. Notably, an increase in the refractive index often leads to enhanced scattering properties of the particle, which makes it more feasible for the radar to correctly detect the nano-particle. In radar systems, received signals often include both echoes and undesired signals, known as noise. A simple method to mitigate noise effects is through \textit{threshold detection}. This involves selecting an appropriate threshold value based on the signal's strength. This threshold helps distinguish between received signals that could be representive of a target's presence and those that are mere noise.
     According to Figure \ref{Mie and RGD results 1}, only specific refractive indices within certain scattering ranges exhibit a noticeable distinction between potential system noise and a valid echo signal. However, it is crucial to note that the effectiveness of the processing unit holds significant importance. This unit is responsible for carrying out subsequent analysis of received echo signals, which includes tasks like determining a reliable threshold value, detecting the presence of a target, and even extracting its electrical properties, such as permittivity. As an instance, in Figure \ref{NsR configuration}, determining an appropriate threshold value aligned with the processing unit's capabilities provides us with the range of detectable scattering angles, denoted as $2\delta$.
%

\begin{table*}
\centering
\caption{Comparison Between RGD and Mie Theorems \cite{Mie calculation 1,Mie calculation 2,Mie Theorem _ 1,Mie Theorem _ 2,RGD Theorem _ 1,RGD Theorem _ 2}}
\begin{tabular}
{|c||c||c|}
\hline
\textbf{Aspect} & \textbf{Rayleigh-Gans-Debye Theory} & \textbf{Mie Theory} \\
\hline
\textit{Particle Size} & Suitable for $a \ll \lambda$ & Applicable to a wide range of particle sizes \\
\hline
\textit{Scattering Regime} & Rayleigh and some Mie scattering & Mie scattering (covers broader scattering range) \\
\hline
\textit{Polarizability Model} & Isotropic & Complex polarizability based on size, shape, and material properties \\
\hline
\textit{Scattering Amplitude} & Simple expression $f(\theta) \propto \cos^2(\theta)$ & More complex expressions based on Bessel functions \\
\hline
\textit{Computational Cost} & Simplified calculations & High computation costs in terms of time and resources \\
\hline
\textit{Accuracy} & Accurate enough on terms (\ref{RGD_Conditions}) & More accurate on different particles \\
\hline
\textit{Usage Scenarios} & Suitable for quick estimations and simple scattering cases & Preferred for detailed and accurate scattering analyses \\
\hline
\end{tabular}
\label{Comparison Between Mie and RGD}
\end{table*}

    This setup suggests an initial step for a NR system, to detect nano-targets. Yet, radar systems encompass tasks beyond mere detection, including estimating the range between the system and the target. While Mie theory addresses electromagnetic wave-particle interaction, it doesn't directly account for calculating the distance between emitter and target, i.e., solely relying on the Mie theory principles, cannot help us through a complete standardization of nanoradars.
     Hence, to unveil deeper insights into target detection, accounting for the coexistence of factors like particle-emitter distance, necessitates the application of more sophisticated electromagnetic theories and computational techniques, such as Finite-Difference Time-Domain (FDTD) approach.
       In other words, to analyze and calculate the waves that are reflected back from the target, we can apply more sophisticated methods and use the Mie or RGD theories to find the wave properties near the nano-particles. For instance, to study the scattering pattern of a nano-object, we can use the FDTD method to simulate the waves at the target location, considering the case of an inhomogeneous and dispersive medium. Then, we can apply the Mie theory to evaluate the backscattered waves.
     Although NRs are still a young field and lack standardized infrastructure, characterization and fabrication methods, certain potential components of these multipurpose systems can be outlined as follows:

     \subsubsection{Nanoantennas}
     Nanoantennas are foundational components within potential NR systems. Engineered to engage with electromagnetic waves and resonate at distinct frequencies, they facilitate the transmission and reception of signals. In other words, they are responsible for controlling wave transmissions and reflections by emitters and targets, respectively. In a potential bistatic NR (BNR) configuration, two separate nanoantennas would be employed—one for transmitting signals and another for receiving them. On the other hand, in a Monostatic NR (MNR) setup, compact transceivers can be utilized to manage both the transmission and reception of radar signals simultaneously \cite{NsR Transceivers 1,NsR Transceivers 2}. Previous sections have discussed numerous structures that have been proposed for potential use in nanoradar systems. Depending on the specific applications and requirements, any of these structures can be used in a nanoradar system. The choice would depend on factors such as desired performance, operating conditions, and target detection capabilities.

      \subsubsection{Photo-detectors (PDs)}
       One convenient method for capturing back-scattered waves from targets within nanochannels involves the use of photo-detectors. These devices possess the ability to sense and detect electromagnetic waves, particularly in the form of light, by generating an electric current when exposed to it.
       The resulting electric current in the external circuitry of the photo-detector, in response to optical radiations, can be monitored and analyzed. This analysis allows for a better understanding of the intrinsic features of the back-reflected light, serving as indicators of both the physical and electrical properties of nano-objects, such as their sizes, speeds, locations, and permittivities \cite{PD_10}. Photo-detectors are considered as the main components of the photo-receiver inside optical channels, e.g., a nanoradar system, with different performance metrics, such as the quantum efficiency, bandwidth, and compatibility, which are expected to be optimized as a primary step of incorporating the PD into a nanoradar configuration. There are different proposed structures for photo-detectors depending on the application, e.g., resonant cavity enhanced photo-detectors (RCE-PD) \cite{PD_1}, carbon nanotube and nanowire-based photo-detectors \cite{PD_2,PD_3}, low-temperature grown gallium arsenide (LT-GaAs) high-speed photo-detectors \cite{PD_4}, plasmonic photo-detectors \cite{PD_5}, CMOS-integrated waveguide photo-detectors \cite{PD_6}, photomultiplier tubes (PMTs) \cite{PD_7}, image intensifiers (I2) \cite{PD_8}, and organic narrowband photodetectors \cite{PD_9}. The possibility of using each of these structures, along with their operational frequency bands must be proven for NR. For instance, we can calculate the photo-generated current of a RCE-PD structure as a result of an incident power as proposed in \cite{PD_1}, i.e.,
    \begin{align}\label{RCE-PD time current}
    & I_{ph}(t)=\frac{q}{x_{a}+w_{n}+w_{p}}[v_{n}N_{ph}(t)+v_{p}P_{ph}(t)],
    \end{align}
    where $q$, $x_{a}$, $N_{ph}(t)$, $P_{ph}(t)$, $v_{n}$, $v_{p}$, $R_{tot}$, and $C_{d}$ are the total charge, active region width, total photo-generated electron concentration, total photo-generated hole concentration, electrons saturation velocity, hole saturation velocity, total resistor of the structure, and the capacitor of the depletion region, respectively. The total photo-generated electron and hole contentrations in (\ref{RCE-PD time current}) are functions of time, dependent on the intrinsic physical parameters of the photo-detector, the incident power intensity $P_{i}$, the propagation frequency $\nu$, and the Planck's constant $\textstyle{h}$ \cite{PD_1}, i.e.,
        \begin{align}\label{Number of photons and holes}
      & \nonumber \nonumber N_{ph}(t)=\frac{P_{i}}{\textstyle{h}\nu}\{(\mu_{f}^{*}+\mu_{b}^{*})[1-e^{-\alpha_{eff}x_{a}}].[u(t)-u(t-\frac{w_{n}}{v_{n}})]+\\ \nonumber
      &[\mu_{f}^{*}(1-e^{-\alpha_{eff}x_{a}+\alpha_{eff}v_{n}t}-\alpha_{eff}w_{n})+\mu_{b}^{*}(-e^{-\alpha_{eff}x_{a}}+\\ \nonumber
      &e^{\alpha_{eff}v_{n}t-\alpha_{eff}w_{n}})][u(t-\frac{w_{n}}{v_{n}})-u(t-\frac{w_{n}+x_{a}}{v_{n}})]\}, \\
      & \\
      & \nonumber P_{ph}(t)=\frac{P_{i}}{\textstyle{h}\nu}\{(\mu_{f}^{*}+\mu_{b}^{*})[1-e^{-\alpha_{eff}x_{a}}].[u(t)-u(t-\frac{w_{p}}{v_{p}})]+\\ \nonumber
      &[\mu_{f}^{*}(1-e^{-\alpha_{eff}x_{a}+\alpha_{eff}v_{p}t}-\alpha_{eff}w_{p})+\mu_{b}^{*}(-e^{-\alpha_{eff}x_{a}}+\\ \nonumber
      &e^{\alpha_{eff}v_{p}t-\alpha_{eff}w_{p}})][u(t-\frac{w_{p}}{v_{p}})-u(t-\frac{w_{p}+x_{a}}{v_{p}})]\}, \\
    \end{align}
    in which $\alpha_{eff}$ is the ionization factor, $\mu_{f}^{*}=\mu_{f}/(1-exp(-\alpha_{eff}x_{a}))$, and $\mu_{b}^{*}=\mu_{b}/(1-exp(-\alpha_{eff}x_{a}))$, where $\mu_{f}$, and $\mu_{b}$ are the forward quantum efficiency (the ratio between the forward optical power to the total incident power), and the backward quantum efficiency (the ratio between the backward optical power to the total incident power), respectively. 

    It's important to highlight that in the next phase, we can analyze the impact of nano-targets on $I_{ph}(t)$ by employing Deep Learning (DL) models, thereby completing the nanoradar detection process \cite{NsR Processor 1}. Further investigation into feasible photo-detectors is essential, if opted for being employed in a NR system. This choice is dependent on multiple factors, including the channel's length, the background material of the channel (water/air/etc.), the optimal number of PDs for effective detection, and their strategic placement within the channel.
      \subsubsection{Signal Processing and Monitoring}
       The processor would assume the responsibility of analyzing radar echo signals, extracting pertinent data, and formulating decisions predicated on the received signals. This unit could be seamlessly integrated within the radar system or function autonomously, establishing connectivity to the radar through suitable interfaces. Additionally, comprehensive oversight and regulation of all signals within the NR system are imperative across various tiers. The radar system's operation, encompassing parameter configuration, configuration adjustments, and overall performance monitoring, necessitates meticulous supervision, facilitated through software interfaces or internal components and controllers. By leveraging Machine Learning (ML) and Deep Learning (DL) techniques, the complex output information contained within nanoscale signals in the NR system, can be efficiently processed and interpreted. This enables researchers and practitioners to gain deeper insights into the data, enabling more precise monitoring and visualization of targets of interest \cite{NsR Processor 1}. Moreover, ML and DL models contribute to the task of detecting and classifying targets, providing enhanced capabilities in accurate identification and categorization. Therefore, the utilization of ML and DL models in analyzing nanoscale signals offers a practical solution that empowers sophisticated data processing, enabling comprehensive understanding and improved decision-making based on the captured information \cite{NsR Processor 1,NsR Processor 2,NsR Processor 3,NsR Processor 4}.
       The processing unit might be empowered by different components and modules. Some of these are as follows:
       \begin{enumerate}[label=(\alph*)]
         \item \textit{Nano-resonators}\\
         Nanoscale resonators, specifically Metamaterial-based (MtM) structures, have the capability to manipulate electromagnetic waves at unique resonant frequencies, determined by their size, shape, and material characteristics. These resonators can serve as highly sensitive sensors, detecting optical forces exerted by nano-objects like quantum dots or individual molecules. Additionally, they offer potential applications in nano-communication systems as precise frequency references for accurate timekeeping \cite{NsR Resonator 1,NsR Resonator 2}. As a result, these structures are designed to enhance signal processing capabilities and improve sensitivity, making them viable components for a potential Nanoscale Radar (NR) system.
         \item \textit{Nano Integrated Circuits (NICs)}\\
          Nanointegrated Circuits (NICs) possess the potential to play a critical role in the processing of radar signals, encompassing essential functions like modulation, sensing and detection (including single-photon detectors), and filtering. The design of these circuits must align with the specific requirements and functionalities of the NR system. Effective operation at the nanoscale and high frequencies, spanning the terahertz and optical ranges, is a key consideration. NICs can be fabricated using innovative materials and structures such as nanowires, graphene, and carbon nanotubes (CNTs) \cite{NsR NICs 1,NsR NICs 2}.
         \item \textit{Nano-actuators and Sensors}\\
         Nanoscale actuators play a crucial role in detecting external stimuli and enabling the NR system to adjust its parameters in response to the surrounding conditions. These actuators possess the unique advantage of precise manipulation at the nanoscale, making them well-suited for a broad range of applications that require accurate and miniaturized motion. The literature presents various types of nanoscale actuators or nanoelectromechanical systems (NEMS). For instance, Piezoelectric Actuators utilize the piezoelectric effect to enable precise positioning and control applications. Electromagnetic Actuators employ magnetic fields to generate forces, while Thermal Actuators rely on the expansion and contraction properties of materials. Additionally, Shape Memory Alloy Actuators utilize shape memory alloys, capable of reversible changes in shape or length when subjected to temperature variations. These diverse types of actuators offer promising avenues for achieving precise and adaptable motion at the nanoscale \cite{NsR Actuator 1,NsR Actuator 2}.
       \end{enumerate}
      Although all of the mentioned building blocks of the NR system have actual fabrications, the compatibility of these components remains a critical challenge. Significant efforts are required to develop the field of nanoradars towards standardization.
       The standardization process aims to ensure consistency, and efficiency across different nanoradar systems. To achieve this goal, researchers and engineers need to focus on explaining the underlying principles of nanoradars and clarifying the essential building blocks required for their implementation. This includes studying the interactions between nanoradar components such as transceiver antennas, nanointegrated circuits (NICs), nano actuators and sensors, and signal processing and monitoring units. In parallel with understanding the fundamental aspects, the development of standardized fabrication methods is a must. Researchers must explore various techniques and processes that enable the efficient and reliable integration of nanoradar components.
       Standardization will enable the development of more robust and reliable nanoradar systems that are versatile tools which find efficient applications across various fields, including biomedical imaging, disease monitoring, drug-delivery supervision, in-body or external monitoring, and more.
\section{Conclusion} \label{FIVE}
In summary, the successful operation of many cutting-edge technologies relies heavily on well-established infrastructures. Optical antennas, particularly those operating in optical ranges including the terahertz gap, are critical components that require thorough investigation and development to meet the demands of state-of-the-art applications such as 6G, IoT, ISAC, nanoscale radar systems, and medical applications. The increasing need for Tbps data transfer rates in optical and terahertz communication channels calls for the development of optical nanoantennas with high gains and wide bandwidths.
This article aims to investigate the fundamentals and emerging technologies in this field, referencing significant publications and offering researchers a comprehensive perspective on the various aspects and challenges connected to the design and construction of nanoantennas for diverse applications. Additionally, for the first time, this paper introduces the standardization of Nano-scale Radar (NR) systems by presenting a foundational building block and discussing potential analysis techniques like Mie and RGD theories to explore the scattering behavior of nanoparticles.
A practical nanoradar must proficiently detect the presence of nanoparticles and extract their physical or electrical properties within the nanochannel. Illuminating nano-objects is achieved through nanoantennas, and capturing back-scattered waves necessitates a device like a transceiver or a photo-detector. The photo-detector monitors the output current generated by incident waves, aiding in the understanding of various aspects and properties of the incoming light. Finally, the data processing unit, empowered by machine learning or deep learning models, analyzes and extracts these properties.
In conclusion, further research is expected in the near future, focusing on feasible configurations within the aforementioned components, ensuring physical compatibility between different parts, and refining data processing techniques for nanoradars.


\end{document}